\documentclass[notitlepage,superscriptaddress,nofootinbib,twocolumn,floatfix]{revtex4-1}

\usepackage{amsmath,amssymb,bm}
\usepackage{graphicx}
\usepackage{xspace}	
\usepackage{siunitx} 
\usepackage{subcaption,booktabs,multirow}

\usepackage{tikz}
\usetikzlibrary{calc}
\usepackage[colorinlistoftodos]{todonotes}

\AtBeginDocument{
	\heavyrulewidth=.08em
	\lightrulewidth=.05em
	\cmidrulewidth=.03em
	\belowrulesep=.65ex
	\belowbottomsep=0pt
	\aboverulesep=.4ex
	\abovetopsep=0pt
	\cmidrulesep=\doublerulesep
	\cmidrulekern=.5em
	\defaultaddspace=.5em
}

\newcommand{\eg}{e.\,g.\xspace}
\newcommand{\ie}{i.\,e.\xspace}
\newcommand{\Ca}{\ensuremath{^{40}\text{Ca}^{+}}\xspace}

\newcommand{\Tc}{\ensuremath{T_\text{c}}\xspace}
\newcommand{\Rm}{\ensuremath{R_\text{m}}\xspace}

\newcommand{\kB}{\ensuremath{k_\text{B}}\xspace}
\newcommand{\Gh}{\ensuremath{\mathit{\Gamma}_\text{h}}\xspace}
\newcommand{\Vdr}{\ensuremath{U_\text{RF}}\xspace}

\newcommand{\wdr}{\ensuremath{\Omega_\text{RF}}\xspace}

\newcommand{\wz}{\ensuremath{\omega_z}\xspace}

\newcommand{\AYBCO}{\ensuremath{A_\text{YBCO}}\xspace}

\usepackage[colorlinks=true, allcolors=blue]{hyperref}

\begin{document}

\title{Electric field noise in a high-temperature superconducting surface ion trap}

\author{P. C. Holz}
\altaffiliation{Corresponding author}
\email{philip.holz@uibk.ac.at}
\affiliation{Institut f\"ur Experimentalphyik, Technikerstra\ss e 25, A-6020 Innsbruck, Austria}
\affiliation{Alpine Quantum Technologies GmbH, Technikerstrasse 17/1, 6020 Innsbruck, Austria}
\author{K. Lakhmanskiy}
\affiliation{Institut f\"ur Experimentalphyik,
Technikerstra\ss e 25, A-6020 Innsbruck, Austria}
\affiliation{Russian Quantum Center, Bolshoy Blvrd 30, Bld 1, 121205 Moscow, Russia}
\author{D. Rathje}
\affiliation{Institut f\"ur Experimentalphyik,
Technikerstra\ss e 25, A-6020 Innsbruck, Austria}
\author{P. Schindler}
\affiliation{Institut f\"ur Experimentalphyik,
Technikerstra\ss e 25, A-6020 Innsbruck, Austria}
\author{Y. Colombe}
\affiliation{Institut f\"ur Experimentalphyik,
Technikerstra\ss e 25, A-6020 Innsbruck, Austria}
\author{R. Blatt}
\affiliation{Institut f\"ur Experimentalphyik,
Technikerstra\ss e 25, A-6020 Innsbruck, Austria}
\affiliation{Institut f\"ur Quantenoptik und Quanteninformation, \"Osterreichische Akademie der Wissenschaften, Technikerstra\ss e 21\,A, A-6020 Innsbruck, Austria}

\begin{abstract}
Scaling up trapped-ion quantum computers requires new trap materials to be explored. Here, we present experiments with a surface ion trap made from the high-temperature superconductor YBCO, a promising material for future trap designs. We show that voltage noise from superconducting electrode leads is negligible within the sensitivity $S_V=\SI{9e-20}{\volt^2\per\hertz}$ of our setup, and for lead dimensions typical for advanced trap designs. Furthermore, we investigate the frequency and temperature dependence of electric field noise above a YBCO surface. We find a $1/f$ spectral dependence of the noise and a non-trivial temperature dependence, with a plateau in the noise stretching over roughly \SI{60}{\kelvin}. The onset of the plateau coincides with the superconducting transition, indicating a connection between the dominant noise and the YBCO trap material. We exclude the YBCO bulk as origin of the noise and suggest further experiments to decide between the two remaining options explaining the observed temperature dependence: noise screening within the superconducting phase, or surface noise activated by the YBCO bulk through some unknown mechanism.    
\end{abstract}

\maketitle

\section{Introduction}
\label{sec:intro}

Trapped ions are among the most promising platforms for quantum information processing to date, with high gate fidelities and long coherence times. In the past years, the complexity of trapped-ion quantum processors has significantly increased, addressing the challenge of scaling to larger numbers of qubits. As such, microfabricated surface traps with many tens of individual electrodes have been realized \cite{Hug2011,Mon2013,Bro2016,Lek2017}, and optics and electronic devices have been integrated into the traps \cite{Meh2020,Nif2020,Stu2019}. However, the increasing number of trap electrodes and growing lead length will eventually cause lead wires to become a significant source of electric field noise due to their finite resistance (Johnson-Nyquist noise). Electric field noise causes heating of the ion motion, compromising the fidelity of entangling gates \cite{Hal2005,Bal2016,Gae2016,Erh2019}. A related problem is the heat dissipation in high-current carrying electrodes, used to generate local B-field gradients for microwave entangling gates \cite{All2013,Bau2019}. The further advance of trapped-ion devices thus requires new trap materials to be explored \cite{Bro2020}.

An interesting class of novel trap materials are high-temperature superconductors (HTSC). Below their critical temperature, superconductors have vanishing electrical resistance and cancel external magnetic fields (Meissner-Ochsenfeld effect). In addition to addressing electric resistance problems, described above, superconductors could also be employed for the integration of efficient on-chip photo detectors, or shielding of magnetic fields from specific trapping regions \cite{Bro2020,Wan2010-2}. 
Furthermore, HTSC could improve the understanding of electric field noise in ion traps, which is still rather poor \cite{Bro2015,Bro2020}. Recent experiments have established that surface noise is a major source of ion heating \cite{All2011,Hit2012,Dan2014,McC2015,Sed2018}, and that this noise can be material dependent \cite{Sed2018}. Hitherto unexplored materials such as HTSC might lead to new insights.

In this paper, we present a first characterization of a HTSC as trap material. Our study focuses on electric field noise in an ion trap made from YBa$_2$Cu$_3$O$_7$ (YBCO), which has a high critical temperature $\Tc\approx\SI{89}{\kelvin}$. Our trap design incorporates electrode leads with a length of about \SI{5}{\milli\meter} and a width of \SI{10}{\micro\meter}, similar to advanced surface trap designs \cite{Ami2010,Mau2016,Hol2020}. As we previously reported \cite{Lak2019}, these leads can be used to probe the superconducting transition with an ion, sensing their tremendous bulk electric noise in the normal conducting regime. Here, we assess the electric noise below the superconducting transition temperature \Tc, showing negligible noise from the leads within the sensitivity of our setup. Furthermore, we investigate the electric field noise above a YBCO surface exposed to a trapped ion. We find an increase of the noise with temperature for $T<\Tc$, and a plateau region for $T>\Tc$ , where the noise is constant over roughly \SI{60}{\kelvin}. The onset of the plateau coincides with \Tc, indicating a relation between noise and electronic properties of the trap material. This is a unique finding in ion traps and strongly differs from previous observations in surface traps made from conventional superconductors \cite{Wan2010,Chi2014}. We discuss several possible explanations for the observed behavior and suggest experiments to further investigate the origin of the noise.

\section{Trap design and setup}
\label{sec:trap-design}

\begin{figure}[htbp]
    \centering
    \includegraphics[width=0.485\textwidth]{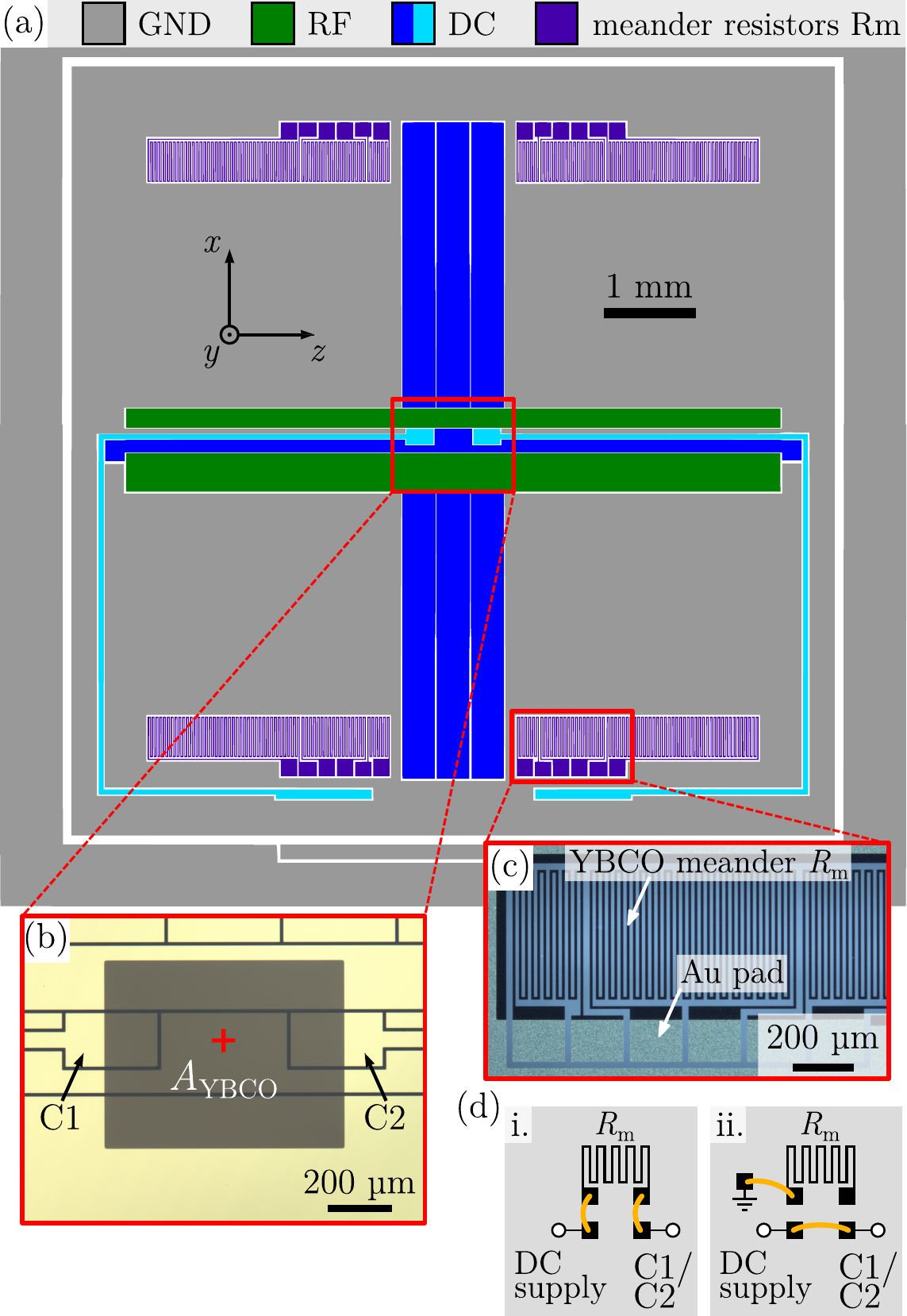}
    \caption{Trap design. (a) Electrode layout with the electrode functionality as color code. (b) Microscope image of the central trapping region. The area \AYBCO with exposed YBCO surface is visible in gray, the trapping position is marked with a red cross. (c) Dark-field image of the YBCO meander leads \Rm (black) with gold bonding pads. (d) Electric configurations. Wire bonds (yellow) are used to connect the electrodes C1 and C2 either through the meander leads \Rm (i.), or directly to their supply lines (ii.).}
\label{fig:trap-layout}
\end{figure}

The ion trap is an adapted form of a standard linear surface trap. An overview of the electrode layout is shown in Fig.\,\ref{fig:trap-layout}\,(a). The trap electrodes are made from \SI{300}{\nano\meter} YBCO on a sapphire substrate, with an additional \SI{200}{\nano\meter} gold layer on top. The gold layer ensures operability of the trap above the critical temperature $\Tc=\SI{89\pm1}{\kelvin}$, where YBCO is a poor conductor. Details on the fabrication are given in Appendix\,\ref{app:YBCO-film}. The center of the trap is shown in Fig.\,\ref{fig:trap-layout}\,(b). Below the trapping position, an area of $\AYBCO=\num{740}\times\SI{580}{\micro\meter^2}$ is not covered by gold, exposing the YBCO surface to a trapped ion. Fig.\,\ref{fig:trap-layout}\,(c) shows the superconducting electrode leads, realized as on-chip meander resistors \Rm and made from YBCO only, \ie without gold top layer. The meanders have a width of $w_\text{m}=\SI{10}{\micro\meter}$ and length of $l_\text{m}=\SI{5.18}{\milli\meter}$, and can be connected to the central DC electrodes C1 and C2 using wire bonds. Identical meander lines at the top of the chip are used to monitor the YBCO film resistance through a 4-wire measurement. 

The trap chip is mounted on a heatable copper carrier inside a cryogenic vacuum apparatus \cite{Nie15}. The heatable carrier, described in appendix~\ref{app:temp-isolation}, is thermally isolated from its cryogenic environment. This allows for trap operation in a wide temperature range $T\in[10,250]\,\si{\kelvin}$, without compromising the cryogenically pumped vacuum or the supply electronics behavior. The trap temperature $T$ can be controlled with an accuracy of about \SI{1}{K}, using the in situ four-wire measurement of the YBCO meander resistors for calibration (more details in \cite{Hol19}). 

Electrically we employ two different configurations for connecting the trap chip to the DC supply, see Fig.\,\ref{fig:trap-layout}\,(d): i. the electrodes C1 and C2 are connected with the supply through the YBCO meander leads. ii. C1 and C2 are directly connected to the supply lines. In either configuration, the other trap electrodes are directly connected to their supply lines, unused meander leads are shorted to the trap's main ground electrode.

\section{Experiment}
\label{sec:experiment}

In the experiments, we confine a single \Ca ion at the trap center, marked with a red cross in Fig.\,\ref{fig:trap-layout}\,(b), at a distance $d = \SI{225}{\micro\meter}$ above the surface. An RF voltage $\Vdr\approx\SI{200}{\volt}$ at $\wdr = 2\pi\times\SI{17.6}{\mega\hertz}$ yields radial frequencies of about $\omega_{x,y}=2\pi\times(2\text{ - }3)\,\si{\mega\hertz}$ (cf. the coordinate system in Fig.\,\ref{fig:trap-layout}\,(a)). The axial frequency is varied within $\wz=2\pi\times(0.4\text{ - }1.8)\,\si{\mega\hertz}$ by scaling the
applied DC voltages. The axial mode is cooled to the ground state using Doppler and sideband laser cooling. The subsequent excitation of the axial motion due to electric field noise is characterized by measuring the ion heating rate \Gh, which is directly proportional to the electric field noise spectral density $S_E$ experienced by the ion, \cite{Bro2015}
\begin{equation}\label{eq:HR}
    \Gh=\frac{q^2}{4m\hbar\wz}S_E(\omega)\,.
\end{equation}
Here $\hbar$ is the reduced Planck constant, $q$ and $m$ are the ion’s charge and mass. The heating rate \Gh is determined using the motional sideband ratio method on the \SI{729}{\nano\meter} $S_{1/2}\leftrightarrow D_{5/2}$ transition  \cite{Lei03}. For each heating rate measurement, typically 5 delay times
are used, each with around 1000 interleaved measurements on the blue and red sideband
transitions. The measurement uncertainties of \Gh are limited by quantum projection noise. 

\begin{figure}[htbp]
    \centering
    \includegraphics[width=0.485\textwidth]{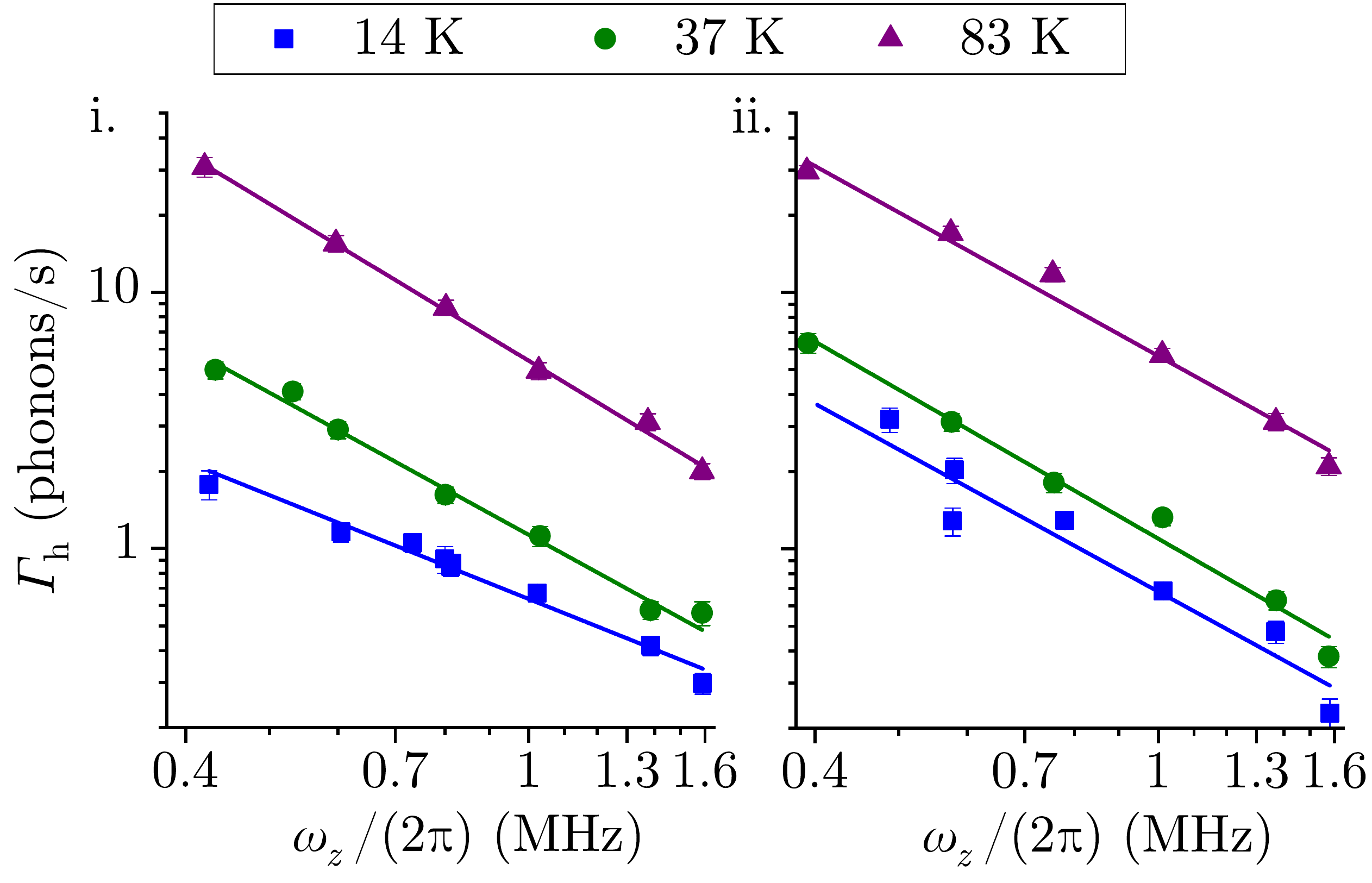}
    \caption{Frequency spectra of the heating rate \Gh for configurations i. and ii., with and without attached YBCO leads, respectively. The three sets are for different temperatures $T<\Tc$, the vertical axes for i. and ii. are identical. Solid lines are fits with a power law, Eq.\,\eqref{eq:power-law}.}
    \label{fig:HR-SC-leads}
\end{figure}

In a first study we investigate the impact of the superconducting electrode leads on the axial heating rate \Gh. In the superconducting state, the meander leads have a negligibly small resistance in the MHz frequency range (see appendix~\ref{sec:app:JN}), and thus produce negligible Johnson-Nyquist noise (JNN). However, \Gh might very well be limited by other bulk noise in YBCO, for instance noise due to the motion of flux vortices \cite{Son1992,Sav1999} or noise related to the presence of grain boundaries \cite{Mar1997,Gus2011}. We thus measure the heating rate in the two different configurations i. (with electrodes C1 and C2 connected through the YBCO meander leads), and ii. (directly connected to the filter lines). The results of these measurements, taken at three temperatures $T<\Tc$, are shown in Fig.\,\ref{fig:HR-SC-leads}. We fit each data set with a power law,
\begin{equation}\label{eq:power-law}
    \Gh(\wz) = \gamma\,(\wz/{\omega_0})^{\alpha-1}\,, 
\end{equation}
where ${\omega_0} = 2\pi\times\SI{1}{\mega\hertz}$. With this definition, $\alpha$ corresponds to the spectral power law scaling of the electric field noise $S_E$. The resulting fit parameters are listed in Tab.\,\ref{tab:PL-fit-results}. The data show nearly perfect agreement between configurations i. and ii. at temperatures $T = \SI{37\pm1}{\kelvin}$ and $T = \SI{83\pm1}{\kelvin}$. At $T = \SI{14\pm1}{\kelvin}$, the data for configuration i. (with YBCO leads) show a slightly lower heating rate at small frequencies $\wz\lesssim 2\pi\times\SI{0.8}{\mega\hertz}$, leading to a smaller value $\alpha=\num{-0.35\pm0.08}$ compared to configuration ii., where $\alpha\approx\num{-0.84\pm0.25}$. We attribute this discrepancy to fluctuations in the measured noise at the lowest measured frequencies, given that the noise magnitude $\gamma$ is almost identical for the two configurations. Overall, the good agreement between configurations i. and ii. at three temperatures $T < \Tc$ excludes that the noise limiting the heating rates in this temperature regime originates in the superconducting meander leads.
\begin{table}[htbp]
	\caption{Power law fit parameters of the data in Fig.\,\ref{fig:HR-SC-leads}.}
	\label{tab:PL-fit-results}
	\centering
	\begin{tabular}[t]{@{} c c c c c c c @{}}
		\toprule
		\multirow{2}{*}{$T$ (K)} & & \multicolumn{2}{c}{$\gamma$ (\si{phonons\per\second})} & & \multicolumn{2}{c}{$\alpha$} \\
		  & & i. & ii. & & i. & ii.\\
		\midrule
		\num{14\pm1} & & \num{0.64 \pm 0.02} & \num{0.68 \pm 0.07} & & \num{-0.35 \pm 0.08} & \num{-0.84 \pm 0.25} \\
		\num{37\pm1} & & \num{1.13 \pm 0.05} & \num{1.10 \pm 0.07} & & \num{-0.85 \pm 0.08} & \num{-0.93 \pm 0.12} \\
		\num{83\pm1} & & \num{5.40 \pm 0.13} & \num{5.65 \pm 0.35} & & \num{-1.04 \pm 0.05} & \num{-0.86 \pm 0.11} \\
		\bottomrule
	\end{tabular}
\end{table}

\begin{figure}[htbp]
    \centering
    \includegraphics[width=0.49\textwidth]{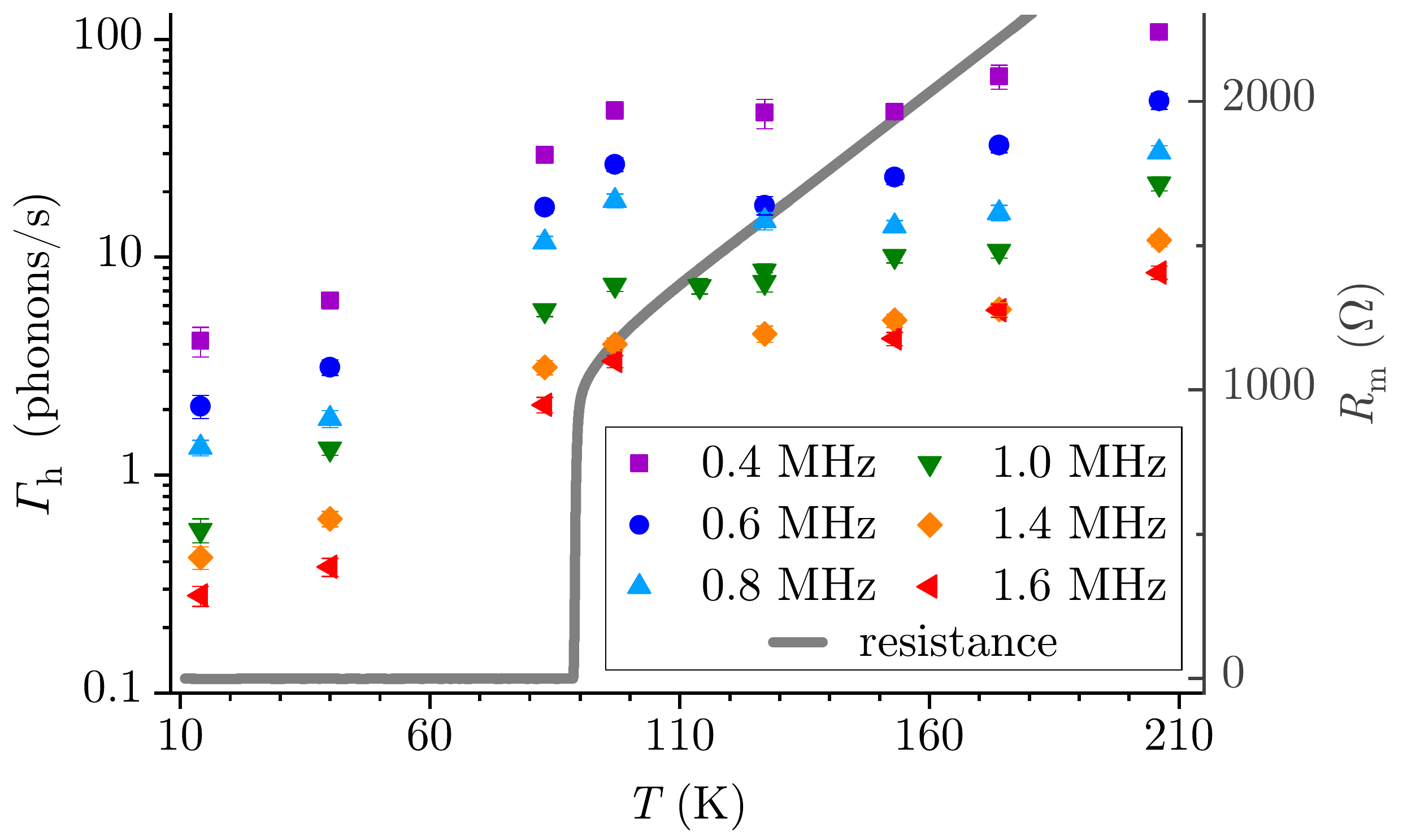}
    \caption{Temperature dependence of the heating rate \Gh for configuration ii., without YBCO electrode leads. The individual sets are taken at different secular frequencies \wz. The gray line (right scale) shows the YBCO meander resistance \Rm, with the critical temperature at $\Tc=\SI{89\pm1}{\kelvin}$.}
    \label{fig:HR-temperature-dependence}
\end{figure}

\begin{figure}[htbp]
    \centering
    \includegraphics[width=0.43\textwidth]{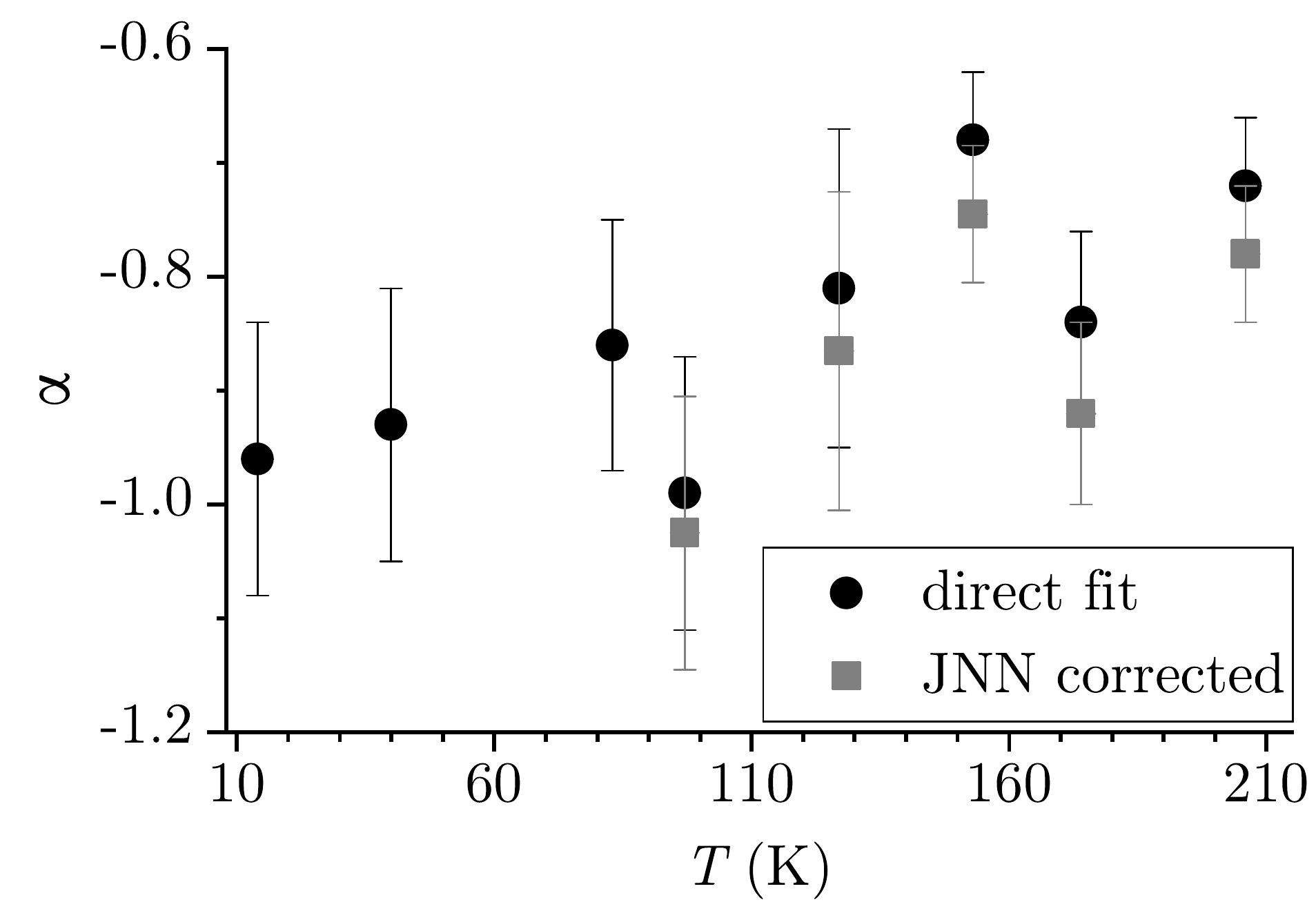}
    \caption{Frequency scaling $\alpha$ of the heating rate data in Fig.~\ref{fig:HR-temperature-dependence}, resulting from a fit with a power law, Eq.\,\eqref{eq:power-law}. The gray squares show the scaling without the contribution of JNN in our setup (see section~\ref{sec:discussion}).}
    \label{fig:HR-PL-exponent}
\end{figure}

In a second study, we further investigate the temperature dependence of the heating rate, by extending the investigated temperature range above the critical temperature \Tc. For these measurements we only use configuration ii., since JNN from the normal-conducting YBCO leads would otherwise dominate \Gh \cite{Lak2019}. The heating rate \Gh in the extended temperature range is shown in Fig.\,\ref{fig:HR-temperature-dependence}. For all axial frequencies \wz, the data show identical behavior:
Below $T\lesssim\SI{90}{\kelvin}$, the heating rate increases strongly with temperature, on average by about a factor 8.2 between $T=\SI{14}{\kelvin}$ and $T=\SI{83}{\kelvin}$. For $T\gtrsim\SI{90}{\kelvin}$ the noise does not follow this trend and \Gh is almost constant over a temperature range of several tens of \si{\kelvin}. The onset of this plateau-like region is around the critical temperature $\Tc=\SI{89\pm1}{\kelvin}$ as evidenced by the in-situ 4-wire resistance measurement (gray data). For temperatures $T\gtrsim\SI{160}{\kelvin}$, the heating rate is slowly rising further, by about a factor 2.5 between $T=\SI{97}{\kelvin}$ and $T=\SI{206}{\kelvin}$. The individual frequency sets are in good agreement with a power law spectrum. Fig.\,\ref{fig:HR-PL-exponent} shows the power law exponent $\alpha$, resulting from a fit to the data in Fig.\,\ref{fig:HR-temperature-dependence} with Eq.\,\eqref{eq:power-law}. The exponent $\alpha$ shows no pronounced difference between the two temperature regimes above and below \Tc, but there appears to be a slight trend with the exponent changing from $\alpha\approx-1.0$ at $T=\SI{14\pm1}{\kelvin}$ to $\alpha\approx-0.7$ at $T=\SI{210\pm1}{\kelvin}$.

Finally, we have measured the heating rate at $\wz=2\pi\times\SI{1.0}{\mega\hertz}$ and $T=\SI{14\pm1}{\kelvin}$ for 6 different axial ion positions in a range $\Delta z\approx\SI{100}{\micro\meter}$ around the trap center. We find a mean heating rate $\Gh=\SI{0.9\pm0.3}{phonons/s}$, without observable trend in the spatial variation.

\section{Discussion}
\label{sec:discussion}

At the coldest temperature, $T=\SI{14\pm1}{\kelvin}$, the measured heating rates show values around $\Gh\approx\SI{1}{phonon/s}$ at $\wz=2\pi\times\SI{1.0}{\mega\hertz}$, equivalent to an electric field noise $S_E\approx\SI{7e-15}{\volt^2\meter^{-2}\hertz^{-1}}$ and comparable to other cryogenic surface traps \cite{Bro2015}. In what follows, we'll refer to this noise level as the background noise in our experiment. The data in Fig.\,\ref{fig:HR-SC-leads} exclude that the background noise is caused by the superconducting electrode leads. Consequently, the intrinsic voltage noise per lead, $S_V^\text{lead}$, must be significantly lower than the background noise, \cite{Bro2015} $S_V^\text{lead}\ll S_E D^2/2\approx\SI{9e-20}{\volt^2\per\hertz}$. Here, we have used the characteristic distance $D=\SI{5.10}{\milli\meter}$ of electrodes C1 and C2, and the factor 2 accounts for the noise from two leads.
Furthermore, it is instructive to compare the background noise with noise that would be produced by electrode leads made from aluminum, but with otherwise identical geometry as the YBCO meander leads. Assuming a typical resistivity of $\rho_\text{Al} = \SI{1e-8}{\ohm\meter}$ at $T = \SI{20}{\kelvin}$ \cite{Cla1970}, such Al leads would induce a JNN limited heating rate of $\Gh^\text{(JNN)}\approx\num{0.2}$\,phonons/s at $\wz=2\pi\times\SI{1.0}{\mega\hertz}$, see Appendix\,\ref{sec:app:JN_Al-lead}. This is only slightly smaller than the background noise in our experiment, illustrating the benefit of superconducting trap materials for advanced trap designs, where a large number of long electrode leads will be required.

In the remainder of this article, we will discuss the possible origin of the background noise, Fig.\,\ref{fig:HR-temperature-dependence}. First, we note that any level of technical noise or JNN passing through the DC filters or RF resonator is practically independent of the trap temperature $T$. This is due to the thermal decoupling between locally heated trap and drive electronics in our setup (see Appendix~\ref{app:temp-isolation} for details). Technical noise from the resistive heater used to increase the trap temperature is ruled out as origin of the observed electric field noise: no change in the measured heating rate was observed upon adding an additional low-pass filter to the heater line ($\approx74\,\si{dB}$ attenuation at 1 MHz). 
Second, we establish that the measured heating rates in Fig.\,\ref{fig:HR-temperature-dependence} do not follow a simple temperature dependence, but are better described by a kink-like dependence with an initial rise, suddenly saturating in a plateau at $T\approx\Tc$. On a purely phenomenological basis, we fit the data both with a simple power law, $\Gh^\text{(1)}(T)$, and with a piece-wise defined power law, $\Gh^\text{(2)}(T)$,
\begin{subequations}\label{eq:PL-temp}
\begin{align}
    \Gh^\text{(1)}(T) &= \varGamma_0\left[1+(T/T_1)^{\beta_1}\right]\\
    \Gh^\text{(2)}(T) &= \begin{cases}
                    \Gh^\text{(1)}(T) &, T<T^*\\
                    \Gh^\text{(1)}(T^*)\left[1+\big((T\!-\!T^*)/T_2\big)^{\beta_2}\right] &, T\geq T^*\,.
                \end{cases}
\end{align}
\end{subequations}
Here, $\varGamma_0(\omega)$ is an independent fit parameter for each of the six frequency data sets, and $T_1$, $\beta_1$, $T_2$, $\beta_2$, $T^*$ are global parameters. The fitted curves are shown in Fig.\,\ref{fig:HR-global-fits}, fit parameters are listed in Tab.\,\ref{tab:T-PL-fit-results} in Appendix~\ref{sec:app:model-comparison}. The simple power law strongly deviates from the measured data around $T\approx\Tc$, at the location of the `kink'. The piece-wise defined power law $\Gh^\text{(2)}$ shows much better agreement, as quantified by the significantly lower Akaike and Bayesian information criteria (AIC,BIC), listed in Tab.\,\ref{tab:AIC-BIC}. Details on the calculation of these criteria is given in Appendix~\ref{sec:app:model-comparison}.
\begin{figure}[htbp]
    \centering
    \includegraphics[width=0.485\textwidth]{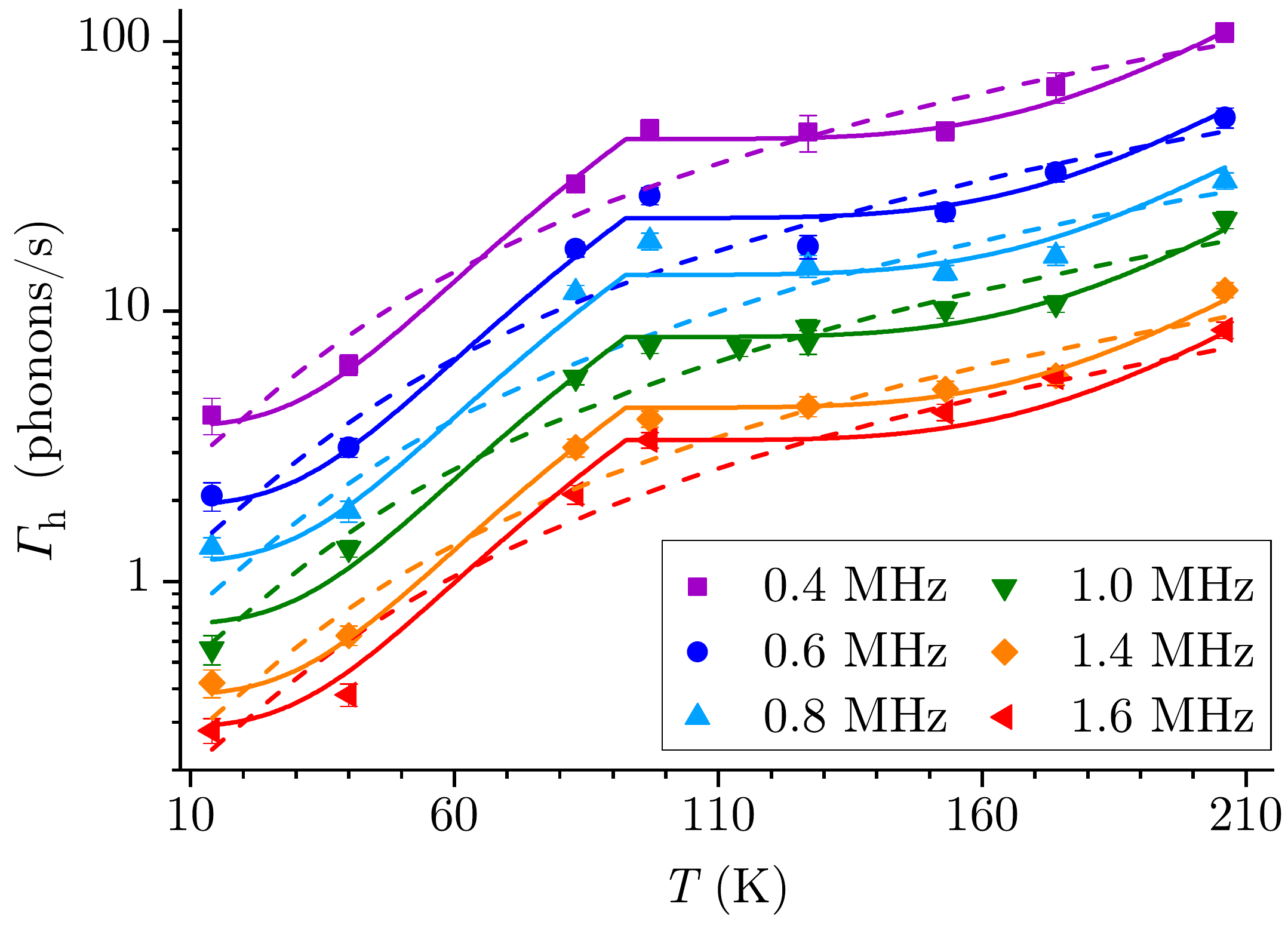}
    \caption{Replot of the heating rate data in Fig.\,\ref{fig:HR-temperature-dependence}, emphasizing the kink-like temperature dependence with a noise plateau at $T\gtrsim\Tc=\SI{89\pm1}{\kelvin}$. A fit with a piece-wise defined power law $\Gh^\text{(2)}(T)$ (solid lines) describes the data much better than a fit with a simple power law $\Gh^\text{(1)}(T)$ (dashed lines).}
    \label{fig:HR-global-fits}
\end{figure}
\begin{table}[htbp]
	\caption{Comparison of the information criteria AIC and BIC of a simple power law model, $\Gh^\text{(1)}(T)$, and a piece-wise defined power law model, $\Gh^\text{(2)}(T)$. $N_\text{p}$ is the number of free model parameters, $N_\text{d}$ the number of data points in Fig.\,\ref{fig:HR-global-fits}. Smaller values of AIC and BIC correspond to a better fitting model.}
	\label{tab:AIC-BIC}
	\centering
	\begin{tabular}[t]{@{} c c c c c c @{}}
		\toprule
		model & & $N_\text{p}$ & $N_\text{d}$ & AIC & BIC \\
		\midrule
		$\Gh^\text{(1)}$ & & 8  & 49 & 136 & 151 \\
		$\Gh^\text{(2)}$ & & 11 & 49 & 55  & 76 \\
		\bottomrule
	\end{tabular}
\end{table}

The fit parameters of the piece-wise defined power law give an onset temperature of the noise plateau $T^*=\SI{92.5\pm2.0}{\kelvin}$, almost identical with the critical temperature $\Tc=\SI{89\pm1}{\kelvin}$ of the YBCO film. Furthermore, we infer the temperature range $\Delta T$ of the plateau, defined as the range where the noise does not increase from its value at $T^*$ by more than the average uncertainty of the heating rate data (about 10\%). The fit parameters give $\Delta T\approx\SI{60}{\kelvin}$.

The clear correlation between the onset of the noise plateau and the superconducting transition, $T^*\approx\Tc$, indicates a relation between the dominant noise and the YBCO trap material. In fact, many physical quantities in high-temperature superconductors show a temperature dependence similar to the dependence of the noise in our trap: the density of superconducting charge carriers \cite{Lan1965}, AC loss \cite{Bha2012}, voltage noise caused by resistance fluctuations \cite{Bar2014}, the magnetic susceptibility \cite{Cha2013}, or the frequency variation of phonon modes \cite{Dri2012}. The fact that there is no marked jump in noise at $T=\Tc$, nor a change in frequency scaling, suggests a single noise source to be dominant over the entire temperature range. Three different scenarios seem likely as an explanation for the observed noise:
\begin{enumerate}
    \item The noise is caused by the YBCO film itself.
    \item The noise is caused by surface noise, activated by processes within the YBCO film.
    \item The temperature dependence of the noise does not stem from the source itself but is introduced via a temperature-dependent attenuation by superconducting screening currents in the YBCO film. 
\end{enumerate}

Scenario 1 is to a great extent ruled out by the measurements in Fig.\,\ref{fig:HR-SC-leads}, which do not show a noise enhancement for attached YBCO leads. Due to the lead geometry, these measurements provide an excellent probe for sources of bulk-noise, \ie sources where the noise increases with the electrode length. Such sources are for instance JNN, but can also have more complex generating mechanisms, \eg related to grain boundaries or flux vortices \cite{Mar1997,Gus2011,Son1992,Sav1999}.
We further rule out noise from the multi-layer trap structure, by means of a noise estimate. For this, we use an approach based on the fluctuation-dissipation theorem (FDT) \cite{Aga1975,Wye1984,Hen1999}, which has been used to predict noise in cold atom and trapped ion experiments \cite{Fer2009,Kum2016,Tel2021}. The FDT formalism assumes an infinite planar trap structure with isotropic layers and neglects electrode gaps. The electric field noise can then be derived from the knowledge of the dissipation of electric energy within the trap material, determined by the material's relative dielectric function $\epsilon(\omega)$ (details in Appendix~\ref{sec:app:BBR}). For YBCO in the superconducting regime, one finds \cite{Fer2009} $\epsilon_\text{YCBO} \approx -\omega^{-2}c^{-2}\lambda^{-2}(T)$, where $c$ is the vacuum speed of light. The temperature dependence of the noise is mainly given by the London penetration depth $\lambda(T) = \lambda_0/\sqrt{1-T/\Tc}$ \cite{Fer2009}, where $\lambda_0\approx(80 \text{ - } 635)\,\si{\nano\meter}$ depending on the YBCO crystal axis \cite{Per2004}. For $T > \Tc$, the YBCO film is treated as a normal metal, as is the Au top layer. The trap's sapphire substrate is modeled as a lossy dielectric.

\begin{figure}[bhtp]
    \centering
    \includegraphics[width=0.485\textwidth]{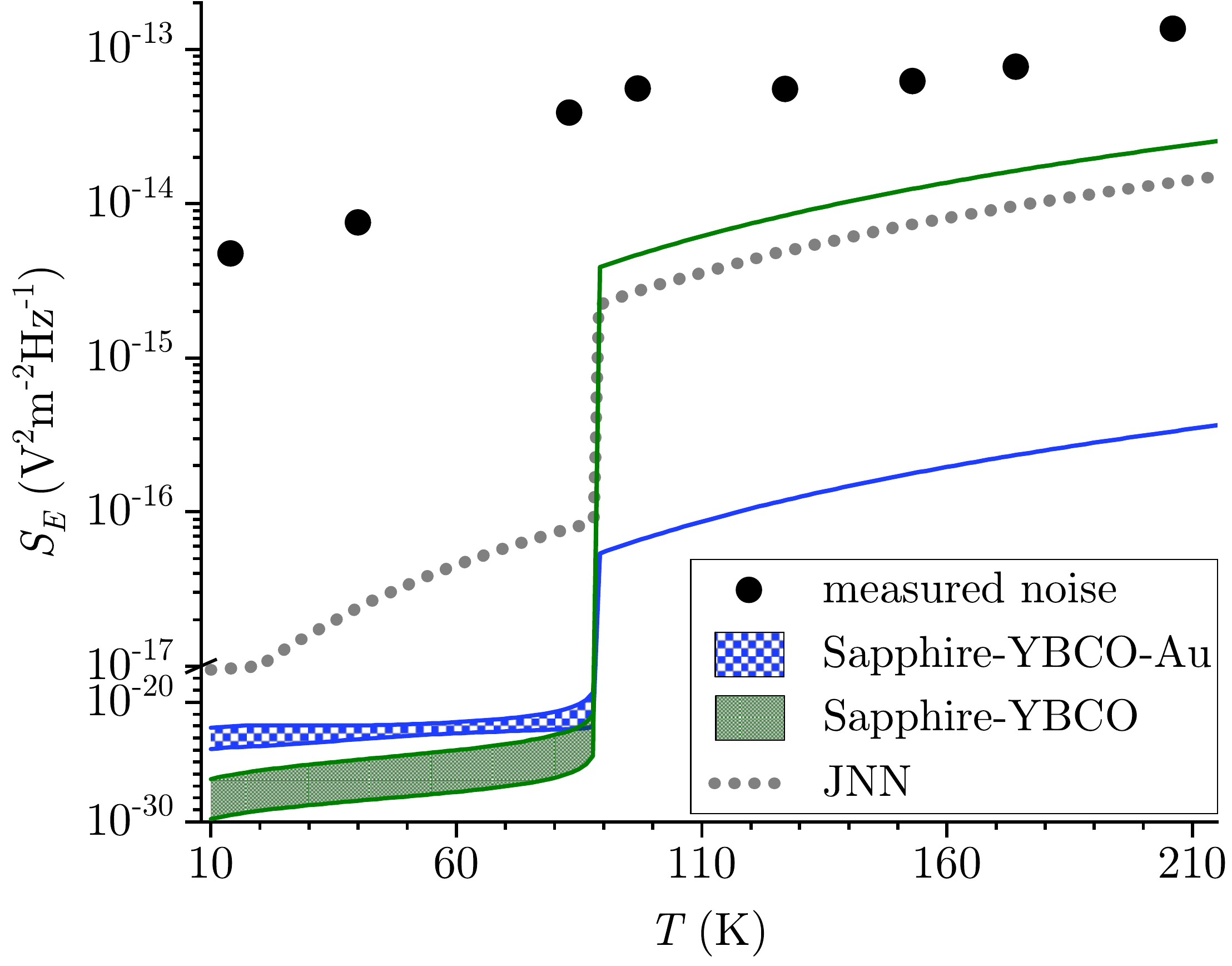}
    \caption{Measured electric field noise and estimates for different noise sources at a frequency $\wz=2\pi\times\SI{1}{\mega\hertz}$. Note the change in the vertical log scale at $S_E=\SI{1e-17}{\volt^2\meter^{-2}\hertz^{-1}}$. The uncertainties in the measured data are smaller than the symbols.}
    \label{fig:noise_models}
\end{figure}

Fig.\,\ref{fig:noise_models} shows FDT estimates together with the measured noise magnitude. The measured data are the power law coefficient $\gamma$ in noise units, derived by fitting Eq.\,\eqref{eq:power-law} to the heating rate data in Fig.\,\ref{fig:HR-temperature-dependence}. The FDT estimates are for a trap layer stack of Sapphire-YBCO-Au (blue), and Sapphire-YBCO (green). The limits of the hatched regions for $T<\Tc$ correspond to the extreme values of the penetration depth literature values, $\lambda_0=\SI{80}{\nano\meter}$ and $\lambda_0=\SI{635}{\nano\meter}$. We note that the noise prediction for the actual trap structure, with the YBCO exposed only at the trap center, should be ranging between the two FDT estimates. Qualitatively, the FDT estimates show a similar temperature scaling for both layer stacks: as the temperature drops below \Tc, the estimates predict a sharp drop of the electric field noise $S_E$ by many orders of magnitude. This is in stark contrast to the kink-like shape of the measured noise. Furthermore, for $T<\Tc$, the predicted noise is negligible compared to the measured noise. For $T>\Tc$ the FDT estimates are closer to the measured noise, but do not reproduce the observed plateau. 

We further note that the temperature dependence of the FDT estimates for $T>\Tc$ is identical to an independent estimate of JNN in our setup, shown as gray dots in Fig.\,\ref{fig:noise_models}. That is, because the JNN estimate is dominated by the electrode resistance in that temperature regime. The difference in noise magnitude stems from the fact that the JNN estimate takes into account the patterning of the trap electrodes, while the FDT estimates neglect the electrode gaps. The JNN estimate also rules out JNN as explanation for the increase in measured noise magnitude at the highest temperatures. For $T>\Tc$, JNN contributes with only 5.3\% to 12.8\% to the measured noise, significantly less than the observed increase in noise by about a factor 2.5. While the contribution of JNN is small, its flat spectral dependence still leads to a small shift of the fitted power law exponent $\alpha$. This shift is shown in Fig.\,\ref{fig:HR-PL-exponent}, where we use the JNN estimate to retrieve the actual exponent $\alpha$ of the observed noise without the JNN contribution (gray squares). Details on the derivation of the JNN estimate are given in Appendix\,\ref{sec:app:JN}.  

The second potential cause of the observed ion heating is surface noise (Scenario 2). Previous studies have observed surface noise with a power-law ($S_E\propto T^\beta$) or Arrhenius ($S_E\propto e^{-T_0/T}$) temperature dependence \cite{Lab2008,Chi2014,Bru2015,Sed2018}. Such dependencies are not compatible with the sudden, kink-like onset of a plateau observed in our data, Fig.\ref{fig:HR-global-fits}. Fitting these models to our data, we rule out both, power-law and Arrhenius scaling, with 99.9\% confidence (details in Appendix~\ref{sec:app:SN}). A levelling-off in the temperature dependence of $E$-field noise, resembling the onset of a plateau, has recently been observed by Noel et al. \cite{Noe2019}, albeit with the levelling-off occurring at a temperature $T\approx\SI{450}{\kelvin}$, much higher than the onset of the plateau in our data. The noise in Ref.~\cite{Noe2019} was shown to be consistent with a thermally activated fluctuator (TAF) model. However, our measured temperature and frequency dependencies, Figs.\,\ref{fig:HR-temperature-dependence} and \ref{fig:HR-PL-exponent}, are not consistent in the context of the TAF model, as we show in Appendix~\ref{sec:app:SN}. 
Putting aside standard surface noise models, one might conceive a simple explanation for the coincidence of the noise plateau onset with the critical temperature, $T^*\approx\Tc$:  the activation of surface noise by a mechanism linked to processes within the YBCO film. Such a hypothetical mechanism could for instance be phononic excitation of two-level fluctuators on the YBCO surface, where the temperature dependence of the phonon mode occupation density is imprinted on the noise. In such a scenario, one would expect noise above the YBCO film to differ from noise above a gold surface. While our measurements at different axial positions did not show an observable spatial variation in noise magnitude, this does not rule out such activation mechanisms: the change in axial position remained small compared to the the extent of the exposed YBCO area $A_\text{YBCO}$ since the trap is designed to create axial confinement only around the center. We note that at the trap center, surface noise from the exposed YBCO surface would dominate with 94\% over surface noise from the surrounding gold area, assuming identical fluctuation strengths on the different surfaces, as we calculate in Appendix\,\ref{sec:app:SN}. 

Third, one might explain the drop in heating rate at $T<\Tc$ with a temperature-dependent attenuation of electric field noise by superconducting screening effects of the YBCO film (scenario 3). In this scenario, the origin of the dominant noise remains unclear and could be either technical noise, JNN or surface noise. For instance, the heating rate might be limited by surface noise with a power-law temperature dependence that is constant below the activation temperature $T_2\approx\SI{200}{\kelvin}$ (see Fig.\,\ref{fig:HR-global-fits}), and which is then screened at $T<\Tc$. Naturally, one would expect the strongest screening for noise sources at the substrate-YBCO interface or within the substrate bulk, similar to effects seen in cold-atoms experiments using superconducting chip traps \cite{Fer2009,Kas2010}. However, considering that the YBCO film imposes boundary conditions on the electric field noise \cite{Hen1999}, it might also influence the noise magnitude at the ion position of more distant noise sources, such as technical noise on one of the trap electrodes. In fact, after heating the trap to $T\approx\SI{250}{\kelvin}$ we have seen a change in the noise level at smaller temperatures, which had previously been reproducible for several temperature cycles of the cryogenic setup. That suggests that the strong local heating may have altered some electrical properties of the supply lines, leading to a change in the observed noise level. This hypothesis could be tested by injecting technical noise through one of the trap electrodes and measuring its dependence on the trap chip temperature. Unfortunately, at the time of the study, such a measurement was not performed.

\section{Conclusion}
\label{sec:conclusion}

In this article we have described the probing of electric field noise in a surface ion trap made from the high-\Tc superconductor YBCO. In our first study we have investigated the impact of superconducting electrode leads on the ion heating rate. The lead dimensions are comparable to those in advanced trap designs that aim at scaling-up trapped-ion quantum computers \cite{Ami2010,Mau2016,Bau2019,Hol2020}. Our data show that bulk noise from the superconducting leads is negligible at a sensitivity-level of $S_V=\SI{9e-20}{\volt^2\per\hertz}$, given by the background noise in our experiment. In comparison, leads made from standard trap materials such as Al create noise similar to this background noise, and would thus lead to excess ion heating in traps with multiple leads as used for ion shuttling. Future trap designs could for instance be realized with a multilayer structure, using a superconducting bottom layer to route supply leads to the trap electrodes on the top layer, which could be made from gold or any other metal.

In the second study we have investigated the possible origin of the background electric field noise in our ion trap. The observed field noise has an approximate $1/f$ frequency scaling and shows a striking temperature dependence, correlating with the superconducting transition: below \Tc, the noise increases with rising temperature, while above \Tc it shows a pronounced plateau, where the noise level is constant over a range of about \SI{60}{\kelvin}. We have ruled out that the noise originates in the bulk of the YBCO film itself, which leaves two options: First, the ion heating is caused by surface noise activated by processes within the YBCO bulk through an unknown mechanism. This mechanism imprints the temperature dependence of the bulk processes onto the measured noise. Second, the measured temperature dependence is caused by a temperature-dependent screening effect in the superconducting phase, attenuating a quasi-constant noise source. This source could be technical noise penetrating through the low-pass filters, but also surface noise with a high activation temperature around $T=\SI{200}{\kelvin}$. Further experiments are required to distinguish between options one and two. Attenuation of external noise sources by a superconducting trap material could be probed by means of technical noise injection \cite{Sed2018-2} and would be of immediate relevance for trapped-ion experiments requiring a low electric-field noise, \eg scalable quantum information processors. Distinguishing between screening of surface noise from the Sapphire-YBCO interface and an activation mechanism of surface noise by processes within the YBCO bulk would be quite challenging. One option might be in-situ cleaning of the chip surface, \eg by ion bombardment \cite{Hit2012,Dan2014}.  A measurement of the spatial variation of the noise above YBCO and metallic surfaces could provide important information as well. Additional insights could be gained by measuring ion heating in several traps with different YBCO material properties, varying the film thickness and oxygen content which controls the charge carrier density \cite{Lia2006}. Such measurements could shed new light on the mechanisms driving surface noise and might even provide input for the understanding of the various phases in YBCO that are still not fully understood \cite{Vis2018}.

\section*{Acknowledgements}
We thank Carsten Henkel for fruitful discussions and the quantum information experiment team for technical assistance.
We acknowledge financial support by the Austrian Science Fund (FWF) through projects P26401 (Q-SAIL) and F4016-N23 (SFBFoQuS),  by  the  Institut  f\"ur  Quanteninformation GmbH, and by the Office of the Director of National Intelligence (ODNI), Intelligence Advanced Research Projects Activity (IARPA), through the Army Research Office Grant No.  W911NF-10-1-0284. This project has received funding from the European Union's Horizon 2020 research and innovation programme under Grant Agreement No.801285 (PIEDMONS). All statements off act, opinion or conclusions contained herein are those of the authors and should not be construed as representing the official views or policies of IARPA, the ODNI, or the U.S. Government.

\bibliography{references}
\bibliographystyle{unsrt}

\clearpage
\section*{Supplemental Information}
\appendix

\section{Fabrication details}
\label{app:YBCO-film}

\begin{figure}[htbp]
    \centering
    \includegraphics[width=0.45\textwidth]{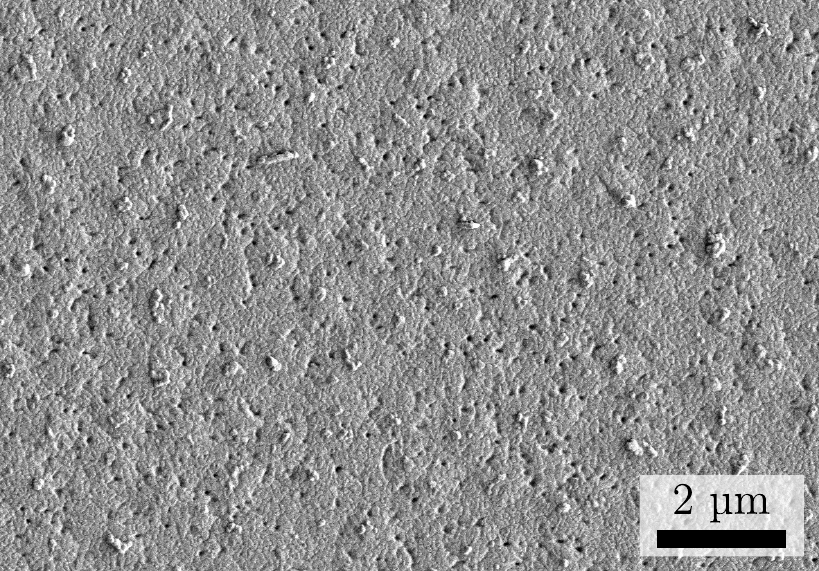}
    \caption{SEM image of the YBCO film produced at ceraco coating GmbH.}
    \label{fig:SEM}
\end{figure}

The trap wafers are coated by ceraco ceramic coating GmbH\footnote{ceraco ceramic coating GmbH, Rote-Kreuz-Str. 8, 85737 Ismaning, Germany}. A YBCO film with thickness $t_\text{YBCO}=\SI{300}{\nano\meter}$ is grown epitaxially on a $\text{CeO}_2$-buffered \SI{500}{\micro\meter} thick sapphire wafer. Figure~\ref{fig:SEM} shows a scanning electrode microscope (SEM) image of the trap's YBCO film. The film has a porous surface with small $\text{CuO}_x$-segregations, a typical appearance for so-called m-type films according to the producer ceraco. The high critical current density $j_\text{c}(\SI{77}{\kelvin})=\SI{3.4}{\mega\ampere\per\centi\meter^2}$, measured at ceraco, indicates the good crystallinity of the film.

Directly after the YBCO film growth, the additional  gold layer with thickness $t_\text{Au}=\SI{200}{\nano\meter}$ is  deposited in the same evaporation chamber. Patterning and dicing of the wafers is done by STAR Cryoelectronics\footnote{STAR Cryoelectronics, 25-A Bisbee Court, Santa Fe, NM 87508-1338, USA}. At the position of the central trapping area \AYBCO, the gold layer is removed by a wet etch process in order to expose the YBCO surface. The same process is used to remove the gold at the YBCO meander leads \Rm. Afterwards the electrodes are patterned by argon ion milling. Before dicing, the chips are coated with a photoresist in order to protect the surface. This resist is removed with acetone in a cleaning step before chip installation.

\section{Temperature isolation}
\label{app:temp-isolation}

\begin{figure}[htbp]
    \centering
    \includegraphics[width=0.485\textwidth]{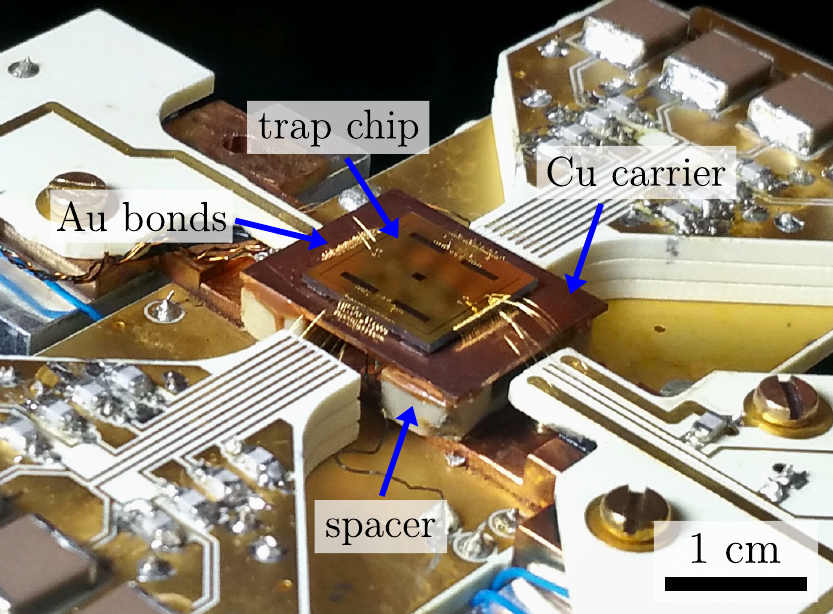}
    \caption{Mounted trap chip with cryogenic electronics PCBs. The trap chip is glued to a copper carrier, thermally isolated from the cold finger by PEEK spacers. Gold wire bonds connect the trap electrodes to the drive electronics. Additional wire bonds at the very edge of the 4 sides of the chip thermally anchor the trap chip to the copper stage.}
    \label{fig:mounted-trap}
\end{figure}

\begin{figure}[htbp]
    \centering
    \includegraphics[width=0.4\textwidth]{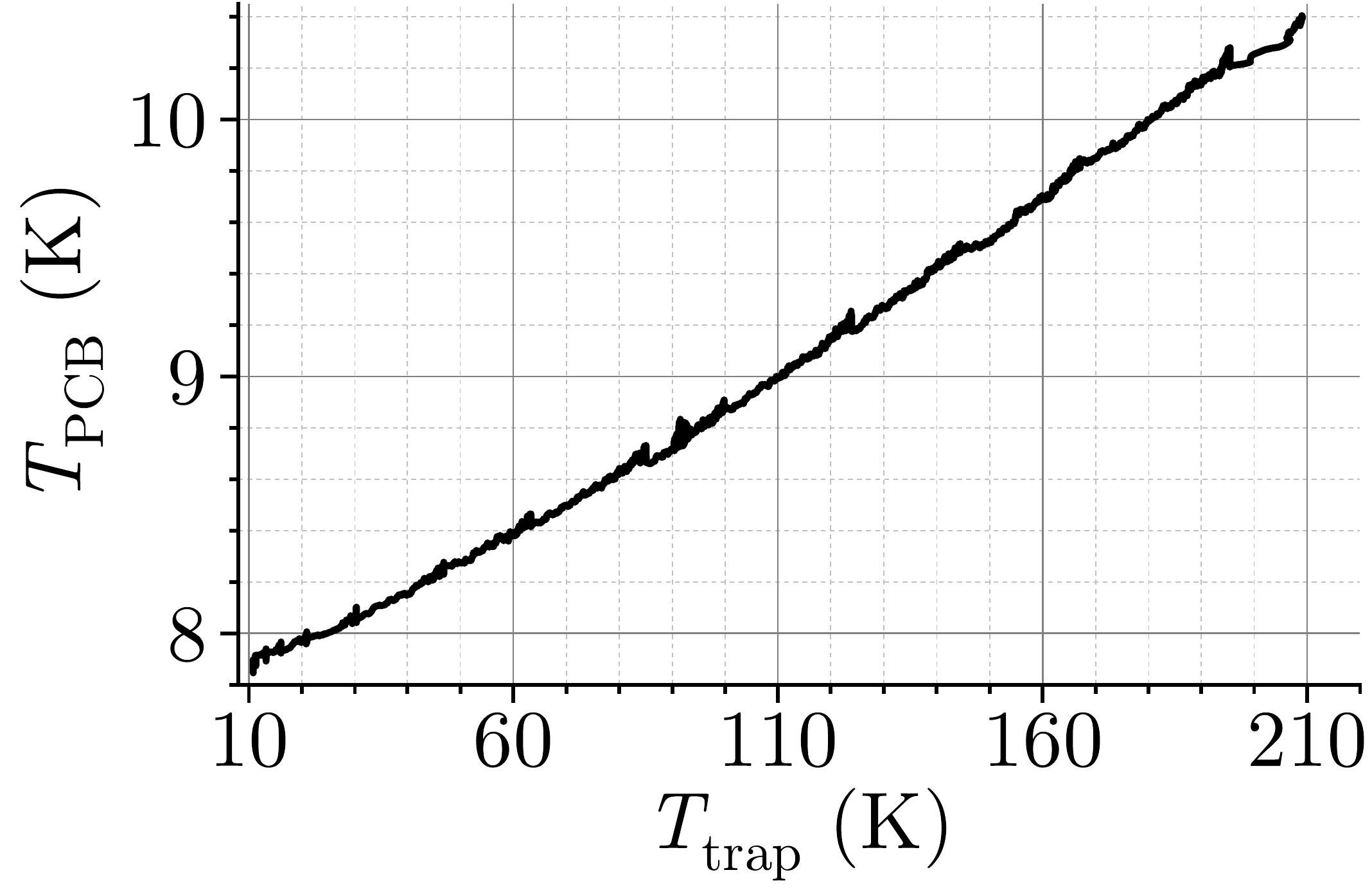}
    \caption{Thermal decoupling between heated trap carrier and surrounding electronics PCBs. The temperatures of heated trap, $T_\text{trap}$, and nearby supply PCBs, $T_\text{PCB}$, are measured with Si diode sensors incorporated in the setup.}
    \label{fig:thermal-insulation}
\end{figure}

The trap is attached to a copper carrier, as shown in Fig.\,\ref{fig:mounted-trap}, which can be resistively heated from below. About 130 wire bonds with a \SI{25}{\micro\meter} diameter gold wire thermally anchor the trap to the copper carrier. The carrier is thermally decoupled from the surrounding PCBs by PEEK spacers. The thermal decoupling reduces the heat load on the cryostat during heating of the trap and helps to maintain the cryogenically pumped vacuum. Furthermore, the nearby supply PCBs with low-pass filters and the RF resonator stay at a nearly constant temperature, as shown in Fig.\,\ref{fig:thermal-insulation}. Heating of the trap from the base temperature $T\equiv T_\text{trap} = \SI{10}{\kelvin}$ to $T_\text{trap} = \SI{210}{\kelvin}$ leads to a temperature increase of the surrounding electronics PCBs from $T_\text{PCB} \approx \SI{8}{\kelvin}$ to only $T_\text{PCB} \approx \SI{10}{\kelvin}$. This thermal decoupling ensures that the low-pass filtering of external electric noise such as technical noise or Johnson-Nyquist noise (JNN) from leads behind the filters is unaffected by a change in trap temperature.  

\section{Phenomenological model comparison for the heating rate temperature dependence}
\label{sec:app:model-comparison}

We establish the kink-like temperature dependence of the heating rate data in Fig.\,\ref{fig:HR-temperature-dependence} by comparing a fit with a simple power law dependence, $\Gh^\text{(1)}(T)$, to a fit with a piece-wise defined power law, $\Gh^\text{(2)}(T)$, cf. Eq.\,\eqref{eq:PL-temp}. The parameters of these fits are listed in Tab.\,\ref{tab:T-PL-fit-results}. From the residual sum of squares (RSS) of the fits we infer the Akaike and Bayesian information criteria (AIC, BIC), listed in Tab.\,\ref{tab:AIC-BIC}, \cite{Ken2004}
\begin{subequations}\label{eq:AIC-BIC}
    \begin{align}
        \text{AIC} &= N \ln(\text{RSS}/N) + 2 M\,,\\
        \text{BIC} &= N \ln(\text{RSS}/N) + M\ln(N)\,,
    \end{align}
\end{subequations}
where $\ln$ is the natural logarithm, $N$ is the number of data points and $M$ is the number of free model parameters. The better fit of the piece-wise power law is clearly quantified by the significantly lower AIC and BIC \cite{Ken2004}.

\begin{table*}[htbp]
	\caption{Fit parameters of the simple power law, $\Gh^\text{(1)}(T)$, and the piece-wise defined power law, $\Gh^\text{(2)}(T)$, in Fig.\,\ref{fig:HR-global-fits}. The parameters $T_1$, $\beta_1$, $T_2$, $\beta_2$, and $T^*$ are global parameters for all frequency data sets.}
	\label{tab:T-PL-fit-results}
	\centering
	\begin{tabular}[t]{@{} c c c c c c c c c c c c c c c c c c @{}}
		\toprule
		\multirow{2}{*}{fit model} & & \multirow{2}{*}{$T_1$ (K)} & & \multirow{2}{*}{$\beta_1$} & & \multirow{2}{*}{$T_2$ (K)} & & \multirow{2}{*}{$\beta_2$} & & \multirow{2}{*}{$T^*$} & & \multicolumn{6}{c}{$\varGamma_0$ (phonons/s)}\\
		 & & & & & & & & & & & & 0.4\,MHz & 0.6\,MHz & 0.8\,MHz & 1.0\,MHz & 1.4\,MHz & 1.8\,MHz\\
		\midrule
		$\Gh^\text{(1)}$ & & \num{22.1 \pm 8.8} & & \num{1.69 \pm 0.14} & & n.a. & & n.a. & & n.a. & & \num{2.18\pm0.95} & \num{1.04 \pm 0.45} & \num{0.62 \pm 0.26} & \num{0.41 \pm 0.17} & \num{0.21 \pm 0.09} & \num{0.16 \pm 0.07} \\
		$\Gh^\text{(2)}$ & & \num{46.2 \pm 3.1} & & \num{3.39 \pm 0.27} & & \num{102.9 \pm 2.8} & & \num{4.14 \pm 0.57} & & \num{92.5\pm2.0} & & \num{3.76 \pm 0.35} & \num{1.92 \pm 0.17} & \num{1.18 \pm 0.10} & \num{0.70 \pm 0.06} & \num{0.38 \pm 0.03} & \num{0.29 \pm 0.03}\\
		\bottomrule
	\end{tabular}
\end{table*}

\section{Noise estimates}
In this appendix we give further details on the electric field noise from different sources: 1. Noise from the trap structure, which is calculated using a fluctuation-electrodynamics approach. 2. JNN from the experimental setup. 3. JNN arising from an electrode lead made from Aluminum. 4. Surface noise.

\subsection{Noise from the trap structure}
\label{sec:app:BBR}

We employ the fluctuation-dissipation theorem to estimate the noise from the multi-layer trap structure \cite{Aga1975,Wye1984,Hen1999}. The formalism assumes an infinite planar trap structure with isotropic layers and neglects electrode gaps. The electric field noise can then be derived from the knowledge of the dissipation of electric energy within the trap material, determined by the material's relative dielectric function $\epsilon(\omega)$. The electric field noise power spectral density $S_E^\text{(trap)}$ of the noise from the trap can be expressed by the free-space black-body radiation $S_\text{BB}(\omega)$ and its modification due to the presence of the trap material, represented by a Green's function $g(\omega)$ \cite{Hen1999}
\begin{equation}\label{eq:FEDN}
    S_E^\text{(trap)}(\omega) = S_\text{BB}(\omega)\left[1+g(\omega)\right]\,.
\end{equation}
In the low-frequency limit, $\hbar\omega\ll \kB T$, valid here, the blackbody radiation can be approximated as \cite{Bro2015}\footnote{We use here the single-sided form for the blackbody radiation spectral density, following Ref.~\cite{Bro2015} rather than \cite{Hen1999}.} $S_\text{BB}(\omega)=2\kB T\omega^2/(3\pi\epsilon_0c^3)$, where \kB is the Boltzmann constant and $\epsilon_0, c$ are the permittivity and speed of light in vacuum. For isotropic media, the Green's function $g_\parallel$ for electric field noise parallel to the trap surface is given by \cite{Hen1999}
\begin{equation}\label{eq:Green-function}
    g_\parallel = \frac{3}{4} \text{Re} \int_0^\infty \frac{u\text{d}u}{v} e^{2i k d v}\left[ R_s(u) + (u^2-1) R_p(u)\right]\,,
\end{equation}
where $v=\sqrt{1-u^2}$, $k=\omega/c$ and $d$ is the ion-surface separation. $R_s$ and $R_p$ are the Fresnel coefficients for $s$- and $p$-polarized waves, respectively. For a simple interface between two media $A$ and $B$ with relative dielectric function $\epsilon_A$ and $\epsilon_B$, the Fresnel coefficients are given by \cite{Wye1984}
\begin{subequations}\label{eq:Fresnel_2layer}
\begin{align}
    R_s^{AB}(u) &= \frac{ \sqrt{\epsilon_A-u^2} - \sqrt{\epsilon_B-u^2} }{ \sqrt{\epsilon_A-u^2} + \sqrt{\epsilon_B-u^2} }\,,\\
    R_p^{AB}(u) &= \frac{ \epsilon_B\sqrt{\epsilon_A-u^2} - \epsilon_A\sqrt{\epsilon_B-u^2} }{ \epsilon_B\sqrt{\epsilon_A-u^2} + \epsilon_A\sqrt{\epsilon_B-u^2} }\,.
\end{align}
\end{subequations}
The dielectric functions $\epsilon$ are in units of the vacuum permittivity $\epsilon_0$. From the basic coefficients, Eq.\,\eqref{eq:Fresnel_2layer}, the Fresnel coefficients of more complicated geometries may be derived, for instance by using a recursion relation \cite{Tom2010}. For a three-layer geometry of medium $A$ (bulk), medium $B$ with thickness $t_B$ and medium $C$ (bulk), one finds \cite{Tom2010}
\begin{equation}\label{eq:Fresnel_3layer}
    R_{s,p}^{ABC}(u) =  \frac{ R_{s,p}^{AB} + R_{s,p}^{BC} e^{2iW_B t_B} }{ 1 - R_{s,p}^{BA} R_{s,p}^{BC} e^{2iW_B t_B} }\,,
\end{equation}
with $W_B=\sqrt{\epsilon_B-u^2}$ \cite{Wye1984}. Changing the volume of medium $C$ from bulk to finite thickness $t_C$, and adding another medium $D$ (bulk) gives rise to a four-layer geometry with \cite{Tom2010}
\begin{subequations}\label{eq:Fresnel_4layer}
\begin{align}
    R_{s,p}^{ABCD}(u) &=  \frac{ R_{s,p}^{ABC} + \mathcal{A}_{s,p}^{ABC}R_{s,p}^{CD} e^{2iW_C t_C} }{ 1 - R_{s,p}^{CBA} R_{s,p}^{CD} e^{2iW_C t_C} }\,,\\
    \mathcal{A}_{s,p}^{ABC} &= \frac{ e^{2iW_B t_B} - R_{s,p}^{AB} R_{s,p}^{CB}}{ 1 - R_{s,p}^{BA} R_{s,p}^{BC} e^{2iW_B t_B} }\,.
\end{align}
\end{subequations}

The Fresnel coefficients $R_s, R_p$ depend on the relative dielectric functions $\epsilon(\omega)$ of the different trap layers: gold, YBCO, and sapphire. For the gold top layer we assume a metallic, conductivity-dominated relative dielectric function \cite{Hen1999}, $\epsilon_\text{Au}=i\sigma_\text{Au}/(\omega\epsilon_0)$, where $\sigma_\text{Au}(T)$ is the electric conductivity taken from \cite{Mat79}. For YBCO, the relative dielectric function for $T<\Tc$ is described within the two-fluid model as \cite{Fer2009}
\begin{equation}\label{eq:epsilon_YBCO}
    \epsilon_\text{YCBO} =  1-\frac{1}{k^2\lambda^2(T)} + \frac{i\sigma_\text{YCBO}(T)}{\omega\epsilon_0}\,,
\end{equation}
with $k=\omega/c$, and $c$, $\epsilon_0$ being the speed of light and electric permittivity in vacuum. The second term in Eq.\eqref{eq:epsilon_YBCO} describes the permittivity contribution due to superconducting electrons, with the London penetration depth $\lambda(T) = \lambda_0/\sqrt{1-T/\Tc}$ following the scaling of $d$-wave superconductors \cite{Fer2009}. For YBCO, the zero-temperature London depth ranges between $\lambda_0\approx(80 \text{ - } 635)\,\si{\nano\meter}$, depending on crystal axis \cite{Per2004}, yielding a large negative relative dielectric function $\epsilon_\text{YCBO}\approx -k^{-2}\lambda^{-2}(T)$ in the regime $T<\Tc$. The last term in Eq.\,\eqref{eq:epsilon_YBCO} describes the contribution of normal conducting electrons, with conductivity $\sigma_\text{YCBO}(T)=\sigma_\text{n} T/\Tc$ and $\sigma_\text{n}$ being the metal conductivity in the normal conducting state \cite{Fer2009}. The conductivity of the normal conducting electrons is given by the value at the transition temperature, $\sigma_\text{n}(T \xrightarrow{>}\Tc)\approx\SI{1.81e6}{\siemens\per\meter}$, extracted from the measured meander lead resistance \Rm. For $T>\Tc$, the YBCO film is treated as a normal metal with $\epsilon_\text{YCBO} = i\sigma_\text{YBCO}/(\omega\epsilon_0)$, where $\sigma_\text{YBCO}(T)$ is again extracted from the measured meander lead resistance \Rm. The trap's sapphire substrate is modeled as $\epsilon_\text{S}=\epsilon_\text{r}(1+i\tan\delta)$, with relative permittivity $\epsilon_\text{r}\approx10$ and loss tangent $\tan\delta=\num{1e-6}$ as specified by the trap manufacturer, ceraco coating GmbH\footnote{The loss tangent is specified at a frequency $\omega=2\pi\times\SI{1}{\giga\hertz}$. A similarly low value $\tan\delta\sim\num{1e-6}$ of single crystal aluminium oxide has been measured in \cite{Vil1998}, without observing significant variations in the frequency range \SI{1}{\kilo\hertz} to \SI{10}{\giga\hertz}.}.

For the noise estimate of the trap region with gold top layer (labelled `Sapphire-YBCO-Au' in Fig.\,\ref{fig:noise_models}) we use the four-layer Fresnel coefficients $R_{s,p}^{ABCD}$ given in Eq.\,\eqref{eq:Fresnel_4layer}, with the following layer assignment: $A\to$ vacuum ($\epsilon(\omega)=1$), $B\to$ Au with thickness $t_B=t_\text{Au}$, $C\to$ YBCO with thickness $t_C=t_\text{YBCO}$, $D\to$ Sapphire. For the noise estimate from the trap region with exposed YBCO (labelled `Sapphire-YBCO' in Fig.\,\ref{fig:noise_models}), we use the three-layer Fresnel coefficients $R_{s,p}^{ABC}$ given in Eq.\,\eqref{eq:Fresnel_3layer}, with layer assignment: $A\to$ vacuum, $B\to$ YBCO with thickness $t_B=t_\text{YBCO}$, $C\to$ Sapphire.

\subsection{Johnson-Nyquist noise}
\label{sec:app:JN}
We estimate the contribution $S_E^\text{(JNN)}$ of JNN to the electric field noise noise experienced by the ion. As sources of JNN we consider the effective low-pass filter resistance $R_\text{filter}$, the resistance $R_\text{lead}$ of the lead wires between the filters and the trap, and the resistance $R_\text{elec}$ of the trap electrodes. We do not consider noise from elements at the \emph{input} side of the low-pass filters since such noise does not depend on the trap temperature $T$, leading to at most a constant noise offset (cf. Appendix~\ref{app:temp-isolation}). The total effective resistance $R_{\text{eff},i}(T) = R_\text{filter}+R_\text{lead}+R_\text{elec}$ is estimated for different temperatures $T\in[10,210]\,\si{\kelvin}$ and for each trap electrode $i$. The resulting electric field noise is then calculated as \cite{Bro2015}
\begin{equation}\label{eq:JN}
    S_E^\text{(JNN)}(T) = 4 \kB T \sum_i \frac{R_{\text{eff},i}(T)}{D_i^2}\,,    
\end{equation}
where \kB is the Boltzmann constant, $D_i$ is the respective electrode's characteristic distance inferred from trap simulation, and the sum runs over all electrodes. 
Eq.\,\eqref{eq:JN} implies that noise from the electrode resistances $R_\text{elec}$ is treated as common noise on the respective electrode, just as $R_\text{lead}$ and $R_\text{filter}$. While such a treatment is in general not valid for $R_\text{elec}$, it is a reasonable approximation in the present trap geometry: For $T<\Tc$, JNN due to electrode resistance is negligible due to the extremely small AC resistivity of YBCO (see below). For $T>\Tc$, about $99.5\%$ of the JNN due to electrode resistance stems from the central electrodes C1, C2 and CC. For these electrodes, the spatial area which mainly adds to the electrode resistance can be separated from the area that dominantly contributes to the characteristic distance $D$, as explained in detail at the end of this section. We now turn to the discussion of the individual contributions to the total effective resistance $R_{\text{eff},i}(T)$.

The low-pass filters are first-order $RC$ with $R=\SI{100}{\ohm}$\footnote{Vishay, Y1625100R000Q9R} and two filter capacitors $C_a=\SI{330}{\nano\farad}$\footnote{Kemet, C2220C334J1GACTU}, $C_b=\SI{470}{\pico\farad}$\footnote{Kemet, C0805C471J1GACTU} in parallel. The effective resistance seen by the ion is \cite{Bro2015} $R_\text{filter} = \text{Re} Z_\text{tot}$, with $Z_\text{tot}^{-1} = R^{-1} + \Sigma_{k=a,b}(\text{ESR}_k+1/(i\omega C_k))^{-1}$. Here, $\text{ESR}_a = \SI{24}{\milli\ohm}$ and $\text{ESR}_b = \SI{1.3}{\ohm}$ are the specified equivalent series resistances of the filter capacitors, which we assume to be temperature independent. Additional filtering due to the trap electrodes' capacitance to ground, which is on the order of \SI{1}{\pico\farad}, is negligible. At a frequency $\omega=2\pi\times\SI{1}{\mega\hertz}$ we find $R_\text{filter}=\SI{26.25}{\milli\ohm}$, limited by $\text{ESR}_a$. 

The lead resistance $R_\text{lead} = R_\text{PCB} + R_\text{bond} + R_\text{contact}$ consists of the trace resistance $R_\text{PCB}$ of the filter PCB (printed circuit board), the resistance $R_\text{bond}$ of the wire bonds to the trap, and the contact resistance between the bonds and the PCB and trap, respectively. The typical contact resistance of a single wire bond produced by our bonding machine to either trap or PCB is $R_\text{contact}=\SI{75}{\milli\ohm}$. The PCB traces have a width $w_\text{tr} = \SI{300}{\micro\meter}$, a thickness $t_\text{tr} = \SI{100}{\micro\meter}$ and an approximate length $l_\text{tr} \approx \SI{1}{\centi\meter}$. The wire bonds have a diameter $D_\text{wb} = \SI{25}{\micro\meter}$ and a typical length $l_\text{wb} \approx \SI{1}{\centi\meter}$. The PCB trace and wire bond resistances result from their respective geometry taking into account the reduction in effective cross sectional area due to the skin effect. For the electric resistivities of the trace (Cu) and the bond (Au) we use the values in \cite{Mat79}. All electrodes except C1 and C2 are doubly bonded which halves their bond and contact resistance. At a frequency $\omega=2\pi\times\SI{1}{\mega\hertz}$, we find the lead resistance of the individual electrodes in a range $R_\text{lead}\sim(50 \text{-} 400)\,\si{\milli\ohm}$ for $T\in[10,210]\,\si{\kelvin}$, limited at low temperatures by the contact resistance and at high temperatures by the wire bond resistance.

\begin{figure}[htbp]
    \centering
    \includegraphics[width=0.485\textwidth]{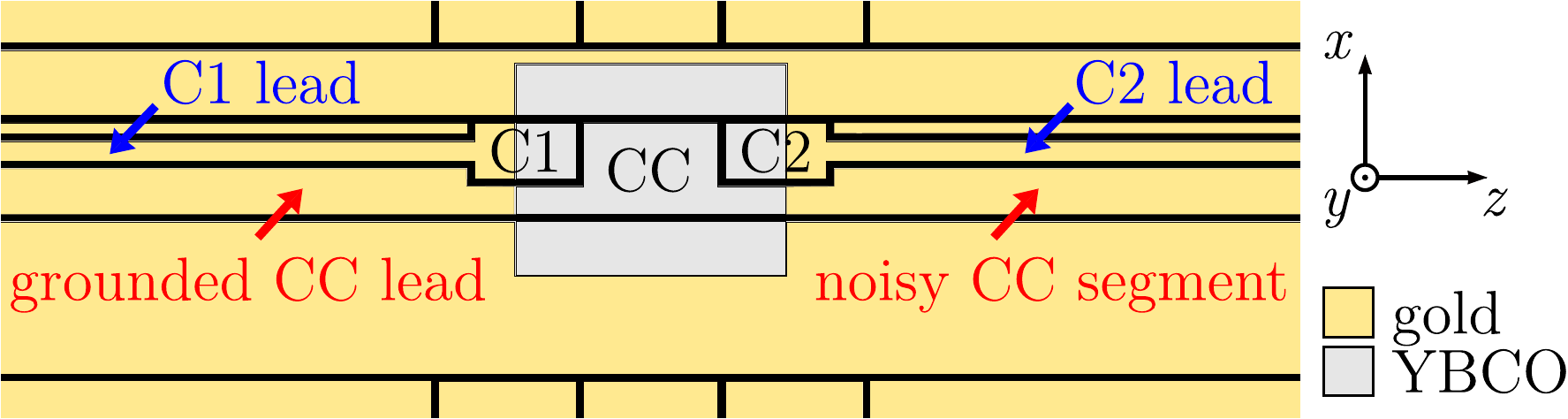}
    \caption{Schematic view of the electrodes in the trap center illustrating the validity of Eq.\,\eqref{eq:JN} for the calculation of JNN due to electrode resistance at $T > \Tc$.}
    \label{fig:JN-electrode-model}
\end{figure}

JNN due to electrode resistance $R_\text{elec}$ is treated in the same fashion as $R_\text{filter}$ and $R_\text{lead}$, \ie as common noise on the respective electrode translating to electric field noise via Eq.\,\eqref{eq:JN}. At temperatures $T>\Tc$, such a treatment is a reasonable approximation for the JNN from electrodes C1, C2 and CC, where we estimate about 99.5\% of the total JNN from electrode resistance to be generated. For electrodes C1 and C2, the dominant resistance stems from the on-chip leads that stretch to the trap edge (cf. Figs.\,\ref{fig:JN-electrode-model} and \ref{fig:trap-layout}), while the characteristic distance $D=\SI{5.10}{\milli\meter}$ practically entirely results from the rectangular area closest to the trap center (labelled `C1',`C2' in Fig.\,\ref{fig:JN-electrode-model}). Electrode CC is connected to the RF ground on the filter PCB only via its left lead (cf. Fig.\,\ref{fig:JN-electrode-model}). The central area of CC, where the poorly conducting YBCO is exposed, thus acts as a resistor adding JNN to the “noisy” Au segment on the right. The relevant characteristic distance $D=\SI{24.3}{\milli\meter}$ of CC is given by the area of the noisy segment.  

The actual electrode resistance $R_\text{elec}$ is calculated from the parallel sheet resistance $R_\text{s}$ of the YBCO and Au films and the electrode geometry. For $T>\Tc$, we determine the YBCO sheet resistance $R_\text{s,YBCO}$ from the 4-wire measurement of the YBCO meander resistance and the meander geometry. The sheet resistance of the Au film is calculated as $R_\text{s,Au}=\rho_\text{Au}/t_\text{Au}$, where $\rho_\text{Au}$ is the Au resistivity from \cite{Mat79} and $t_\text{Au}=\SI{200}{\nano\meter}$ is the Au film thickness. The skin effect can be neglected since the skin depth in both YBCO and Au is much larger than the respective film thickness. The resulting parallel sheet resistance of the YBCO and Au films is dominated by the Au sheet resistance, due to the poor conductivity of YBCO in the normal conducting regime: at $T=\SI{90}{\kelvin}$, we calculate $R_\text{s,YBCO}\approx\SI{1.8}{\ohm}$, while $R_\text{s,Au}\approx\SI{28}{\milli\ohm}$. The electrode resistance follows from the trap geometry treating each electrode as a wire, $R_\text{elec} = R_\text{s} l_\text{s} /w_\text{s} $, where $l_\text{s},w_\text{s}$ are the length and width of the respective electrode.
For $T\in[90,210]\,\si{\kelvin}$, we find an electrode resistance in the range $R_\text{elec}\sim(5.43 \text{ - } 21.3)\,\si{\ohm}$ for electrodes C1, C2 and $R_\text{elec}\sim(11.8 \text{ - } 49.0)\,\si{\ohm}$ for electrode CC. All other electrodes have more than a factor 15 smaller resistances than C1 and C2, and a 3 to 15 times larger $D$ (all characteristic distances are listed in \cite{Hol19}). 

For $T<Tc$, the electrode resistance can be neglected due to the extremely small AC resistivity of YBCO: At $f=\SI{10.9}{\giga\hertz}$ and $T=\SI{10}{\kelvin}$, our YBCO film is specified to have a sheet resistance of about $R_\text{s}=\SI{0.11}{\milli\ohm}$. This is in good agreement with the sheet resistance $R_s=\SI{0.22}{\milli\ohm}$ measured in Ref. \cite{Wan2007} for a \SI{600}{\nano\meter} thick YBCO film, similar to our thickness $t_\text{YBCO}=\SI{300}{\nano\meter}$ (we note that the sheet resistance of the \SI{600}{\nano\meter} film is limited by the penetration depth \cite{Wan2007}). For that \SI{600}{\nano\meter} film, the sheet resistance at $f=\SI{20}{\mega\hertz}$ was measured to be about $R_\text{s}=\SI{1e-7}{\ohm}$ at $f=\SI{20}{\mega\hertz}$ \cite{Wan2007}. This is 5 orders of magnitude below the sheet resistance $R_\text{s,Au}$ of Au, calculated above for $T=\SI{90}{\kelvin}$. The resulting electrode resistances $R_\text{elec}$ for $T<\Tc$ are thus in the $\SI{1e-4}{\ohm}$ range, negligible compared to the lead and filter resistances $R_\text{lead}$ and $R_\text{filter}$.

\subsection{Johnson-Nyquist noise from an Al electrode lead}
\label{sec:app:JN_Al-lead}
We calculate the JNN of an electrode lead with the same geometry as the YBCO leads employed in our experiment, but made from Aluminum. Assuming a resistivity of $\rho_\text{Al} = \SI{1e-8}{\ohm\meter}$ at $T = \SI{20}{\kelvin}$ \cite{Cla1970}, and using the meander geometry given in section\,\ref{sec:trap-design}, we arrive at an Al lead resistance of $R_\text{Al}=\rho_\text{Al} l_\text{m}/(w_\text{m} t_\text{YBCO})=\SI{17.3}{\ohm}$. With the characteristic distance $D=\SI{5.10}{\milli\meter}$ of electrodes C1 (and C2), and employing Eq.\,\eqref{eq:JN}, this leads to an electric field noise of $S_E^\text{(JNN)}=\SI{7.33e-16}{\volt^2\meter^{-2}\hertz^{-1}}$ per Al lead. For two leads, attached to electrodes C1 and C2, we thus arrive at an ion heating rate of $\Gh^\text{(JNN)}=\num{0.21}$\,phonons/s at $\wz=2\pi\times\SI{1.0}{\mega\hertz}$, cf. Eq.\,\eqref{eq:HR}.

\subsection{Surface noise}
\label{sec:app:SN}

\begin{figure}[htbp]
    \centering
    \includegraphics[width=0.485\textwidth]{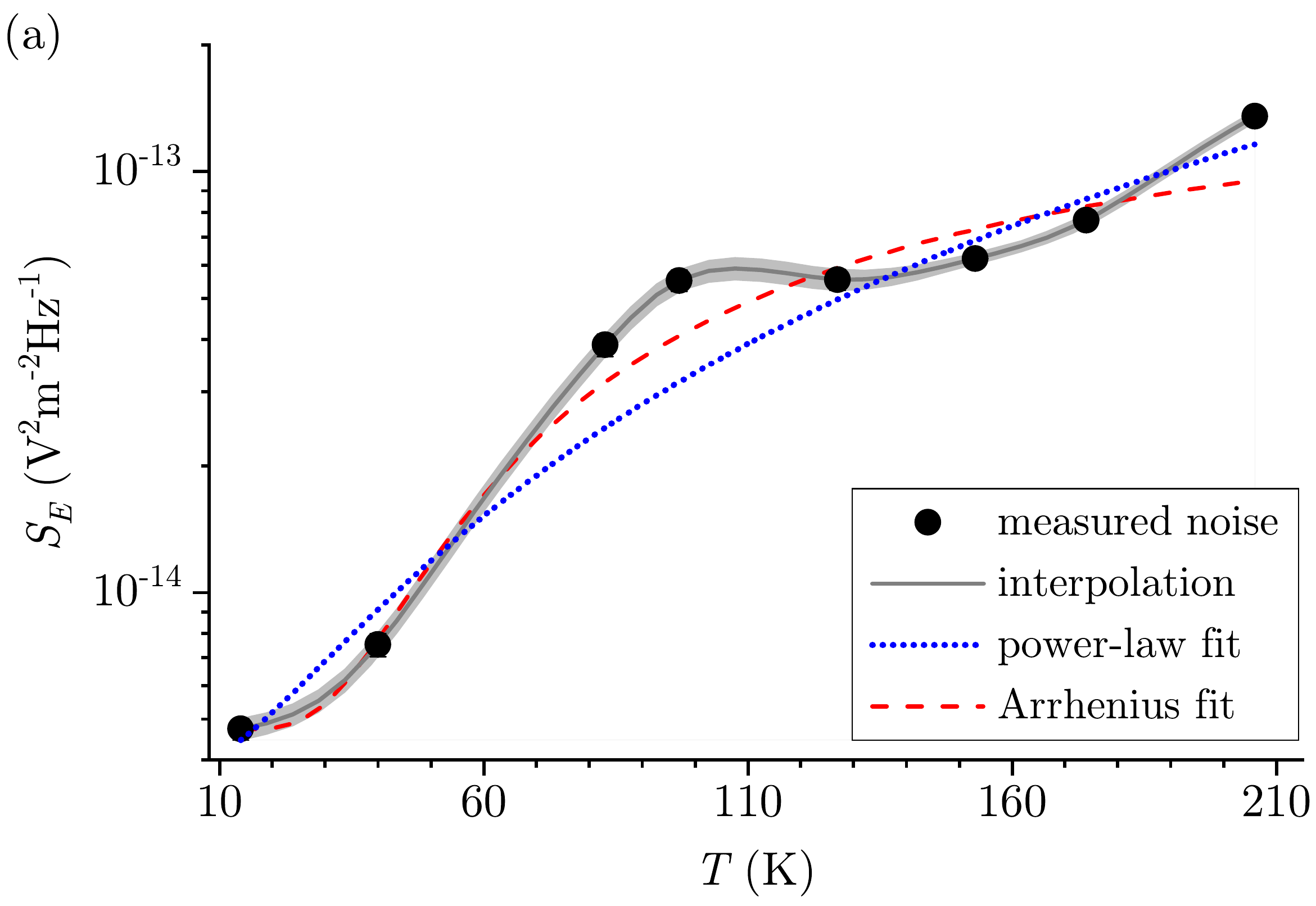}
    \includegraphics[width=0.485\textwidth]{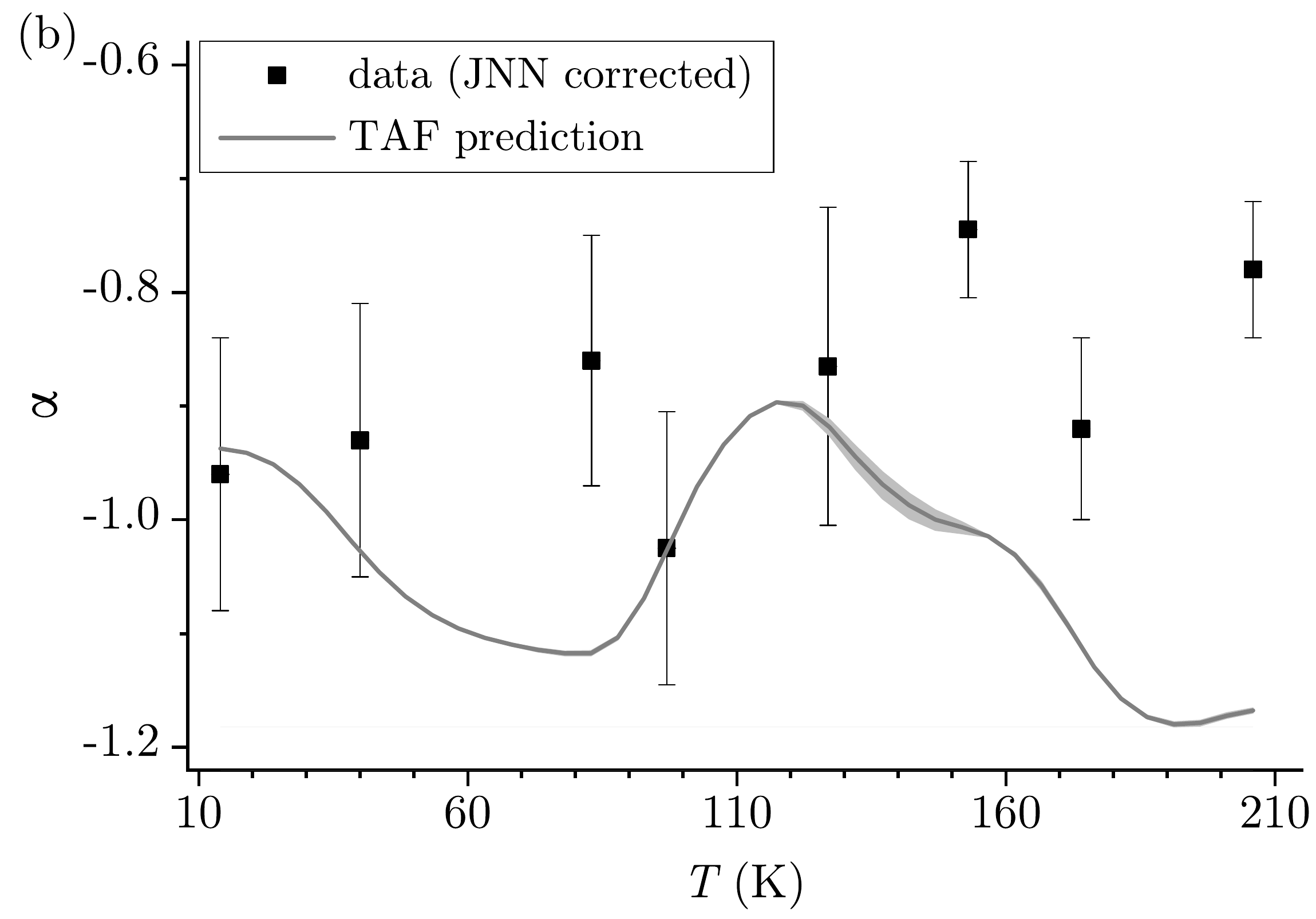}    
    \caption{Disagreement between the measured electric field noise and typical surface noise scalings. (a) Measured noise magnitude and fits with a power-law and Arrhenius scaling. The uncertainties of the measured noise are smaller than the symbols. The gray line is a spline interpolation of the data, the shaded area reflects the confidence interval given by the uncertainty of the measurement. (b) Power-law exponent of the measured frequency scaling and the prediction of the TAF model, based on the spline interpolation in (a).}
    \label{fig:HR_fits}
\end{figure}

We compare the temperature and frequency dependence of the heating rate measurements in Fig.~\ref{fig:HR-temperature-dependence} with scalings associated with surface noise. Starting with the temperature dependence, Fig.\,\ref{fig:HR_fits}\,(a) shows the measured noise magnitude in units of $S_E$, derived  from  fitting the power law, Eq.\,\eqref{eq:power-law}, to the heating rate data in Fig.~\ref{fig:HR-temperature-dependence}. The dotted blue line and dashed red line show fits with a power-law and Arrhenius scaling, respectively. Such scalings are predicted by several surface noise models, and have been observed experimentally \cite{Bro2015,Lab2008,Chi2014,Bru2015,Sed2018}. The power-law fit,  $S_E = S_{E,0}(1+ T/T_0)^\beta$, gives an offset $S_{E,0}=\SI{3.7\pm1.9e-15}{\volt^2\meter^{-2}\hertz^{-1}}$, a temperature exponent $\beta=\num{1.9\pm0.4}$ and characteristic temperature $T_0=\SI{32\pm19}{\kelvin}$. The Arrhenius fit, $S_E= S_{E,0}+ S_{E,T} e^{-T_0/T}$, gives an offset $S_{E,0}=\SI{4.7\pm1.5e-15}{\volt^2\meter^{-2}\hertz^{-1}}$, a value $S_{E,T}=\SI{2.1\pm0.6e-13}{\volt^2\meter^{-2}\hertz^{-1}}$, and a characteristic temperature $T_0=\SI{169\pm43}{\kelvin}$. However, neither scaling fits the data well, as indicated by the reduced chi squared $\chi^2_\text{PL}=29.6$ (power-law) and $\chi^2_\text{A}=31.7$ (Arrhenius), with $p$-values below $10^{-3}$, ruling out these scaling on the $99.9\%$ confidence level. 
An arbitrary temperature dependence can be explained in the context of a thermally activated fluctuator (TAF) model \cite{Noe2019}. The temperature dependence of the noise resulting from an ensemble of fluctuators depends on the density distribution of fluctuator energies. The TAF model predicts that the temperature and frequency scaling are linked, and a local power law exponent $\alpha$ for the spectral dependence can be defined with \cite{Noe2019}
\begin{equation}\label{eq:TAF}
    \alpha(\omega,T) = 1 - \frac{1}{\ln(\omega\tau_0)}\left( \frac{\partial\ln S_E}{\partial\ln T} - 1 \right)  \,,    
\end{equation}
where $\tau_0\approx\SI{1e-13}{\second}$ is a typical hopping attempt time. Using Eq.\,\eqref{eq:TAF} we derive the frequency exponent $\alpha$ predicted by the TAF model from a spline interpolation of the measured electric field noise $S_E$ (solid gray line in Fig.\,\ref{fig:HR_fits}\,(a)). The prediction for $\omega=2\pi\times\SI{1.0}{\mega\hertz}$ is shown in Fig.\,\ref{fig:HR_fits}\,(b), together with the measured frequency exponents resulting from directly fitting the data in Fig.\,\eqref{fig:HR-temperature-dependence} with a power-law, Eq.\,\eqref{eq:power-law}. The TAF model predicts more negative values $\alpha\approx-1.1$, $\alpha\approx-1.2)$ around temperatures $T\approx\SI{80}{\kelvin}$ and $T\approx\SI{200}{\kelvin}$, respectively, where the noise $S_E$ has a strong increase with temperature. This prediction is not well supported by the measured $\alpha$. We quantify the deviation between data and TAF prediction, following the statistical analysis described in Appendic~C of Ref.~\cite{ber2021}. For the data in Fig.\,\ref{fig:HR_fits}\,(b), we calculate a $\chi^2=72.5$ and a $p$-value $p\ll10^{-10}$, showing the inconsistency between measured temperature and frequency dependence within the TAF model. 

Finally, we estimate the contribution of surface noise from the exposed YBCO surface to the total surface noise from both YBCO and Au surfaces at the ion position $\vec{r}_0$. Surface noise $S^\text{SN}_{E,A}$ from a given area $A$ on the chip surface is modeled as the sum of the noise from small patches, $S^\text{SN}_{E,A}\propto\sum_{i\in A}f_i E^2_{z,i}$, where $E_{z,i}$ is the axial component of the field noise produced by patch $i$, $f_i$ is a weighting factor for the intrinsic fluctuation strength on that patch, and the sum runs over all patches within $A$. The noise fraction $\zeta$ caused by patches from within the exposed YBCO area $A_\text{YBCO} = 740\times580\,\si{\micro\meter^2}$ is then given by $\zeta = S^\text{SN}_{E,A_\text{YBCO}}/ S^\text{SN}_{E,A_\text{chip}}$, where $A_\text{chip} = 10\times10\,\si{\milli\meter^2}$ is the approximate total trap chip surface area placed centrally below the trapping position $\vec{r}_0$. The electric field noise contributions from the individual patches at the ion position $\vec{r}_0$ are determined by calculating the axial electric field component $E_{z,i}(\vec{r}_0)$ produced by a constant DC voltage on the respective patch using trap simulation. The patches all have equal dimensions of $1\times1\,\si{\micro\meter^2}$. Convergence of the result is verified by varying the patch size between $0.1\times0.1\,\si{\micro\meter^2}$ and $5\times5\,\si{\micro\meter^2}$. The calculation is additionally validated by confirming the $1/d^4$ scaling of the noise from the entire chip area $S^\text{SN}_{E,A_\text{chip}}$ ($f_i=1\forall i$) with the distance $d$ from the surface, as expected from patch potential noise \cite{Bro2015}. 

\begin{figure}[htbp]
    \centering
    \includegraphics[width=0.4\textwidth]{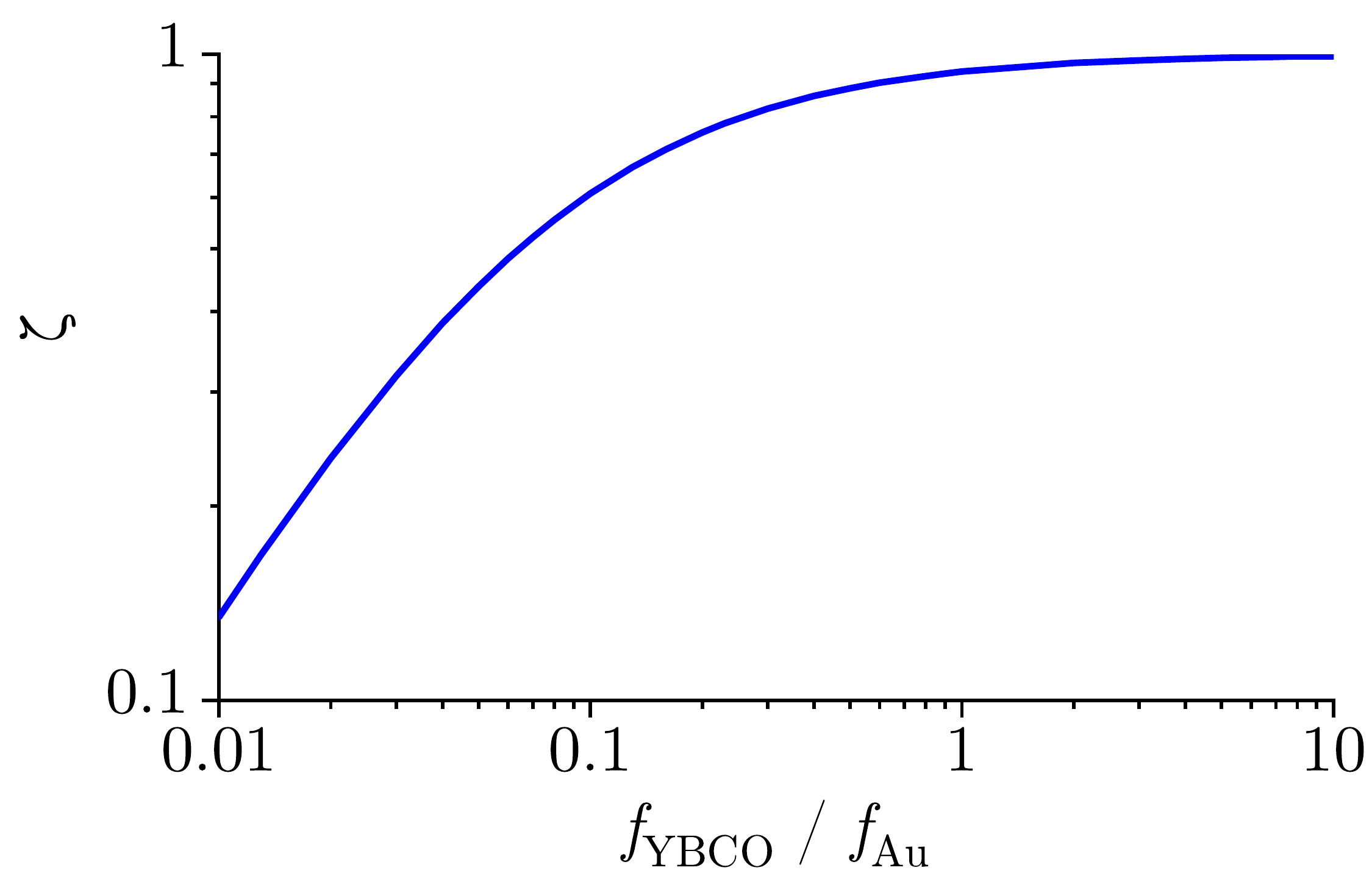}
    \caption{Fraction $\zeta$ of surface noise from the exposed YBCO area to the noise from the entire chip area, as function of the ratio between patch fluctuation strength on YBCO and gold.}
    \label{fig:noise-fraction-YBCO}
\end{figure}

For the calculation of the noise fraction $\zeta$, we assume for simplicity that all patches on the gold surface have equal fluctuation strength $f_\text{Au}$, while patches on the YBCO surface have fluctuation strength $f_\text{YBCO}$. The resulting noise fraction $\zeta$ as function of the ratio  $f_\text{YBCO}/f_\text{Au}$ is shown in Fig.\,\ref{fig:noise-fraction-YBCO}. The data show that noise from the YBCO area dominates over a wide range of fluctuation strength ratios down to approximately $f_\text{YBCO}/f_\text{Au}=0.06$, where $\zeta = 50\%$. At equal fluctuation strength, $f_\text{YBCO}/f_\text{Au}=1$, we find that $\zeta = 93.9\%$.

\end{document}